 \pdfoutput=1
 
\documentclass[12pt]{article}

\usepackage{caption}
\usepackage{subcaption}
\usepackage{graphicx}
\usepackage{graphics}
\usepackage{pifont}
\usepackage{adjustbox}
\graphicspath{{figures/}}
\usepackage{makecell,multirow}
\usepackage{soul}

\usepackage{algorithmic}
\usepackage{array}
\usepackage{listings}
\usepackage{hyperref}
\usepackage{float}
\usepackage{lscape}
\usepackage{xcolor}
\usepackage{enumitem}

\usepackage{mathtools}

\usepackage{amsmath,amsfonts,amssymb,amsthm,epsfig,epstopdf,titling,url,array}

\usepackage[]{apacite}
\definecolor{DarkGreen}{rgb}{0,0.6,0}

\theoremstyle{plain}

\newtheorem*{remark}{Remark}

\theoremstyle{definition}

\theoremstyle{remark}

\begin{document}

\title{Replicating and extending chain-ladder via an age-period-cohort structure on the claim development in a run-off triangle} 


\author{Gabriele Pittarello \thanks{ Sapienza University of Rome
Department: School of Statistical Sciences, gabriele.pittarello@uniroma1.it} \and Munir Hiabu\thanks{University of Copenhagen, Department of Mathematical Sciences, mh@math.ku.dk} \and Andr\'{e}s M. Villegas\thanks{UNSW Sydney, School of Risk and Actuarial Studies, a.villegas@unsw.edu.au}}

\maketitle 

\begin{abstract}
This paper introduces a new stochastic model replicating chain-ladder estimates and furthermore considers extensions that add flexibility to the modeling.
In its simplest form, the proposed model replicates the chain-ladder's development factors using a GLM model with averaged hazard rates running in reversed development time as response.
This is in contrast to the existing reserving literature within the GLM framework where claim amounts are modeled as response. Modeling the averaged hazard rate corresponds to modeling the claim development and is arguably closer to the actual chain-ladder algorithm. 
Furthermore, since exposure does not need to be modeled, the model only has half the number of parameters compared to when modeling the claim amounts. 
This lesser complexity can be used to easily introduce model extensions that may better fit the data. We provide a new R-package, \texttt{clmplus}, where the models are implemented and can be fed with run-off triangles.
We conduct an empirical study on  30 publicly available run-off triangles making a case for the benefit of having \texttt{clmplus} in the actuary's toolbox.
\end{abstract}

\section{Introduction}

{\color{black} The claims reserve is the main liability of a non-life insurer and consists of the insurer's estimate of future payments for claims that have been incurred but not yet paid. Claims reserving is the process of calculating the claims reserve and is traditionally performed by reserving actuaries using simple algorithms. The most commonly used reserving algorithm is the chain ladder, which is based on the run-off triangles of incremental payments, a bivariate aggregation of past claims payments by accident period and development period. This paper presents a new class of models for claims reserving where the simplest form replicates the chain ladder estimates
while the other models may account for additional patterns in the data.
}

Claims reserving models replicating the chain-ladder estimates by modeling the incremental claim amounts have been extensively studied in the literature, see e.g. \citeA{england99}, \citeA{pinheiro03},  \citeA{kuang08b}, \citeA{kuang08}, \citeA{kuang11},
\citeA{bjorkwall09}, \citeA{gabrielli20}, \citeA{avanzi20}, \citeA{taylor21}, {\color{black} \citeA{sriram21}, and \citeA{chang23}}. 
By contrast, given a run-off triangle, we propose to model the claim development instead of the claim amount. It is surprising that despite  modeling the claim development being arguably more in the spirit of the chain-ladder algorithm, there is only little literature that models the claim development given a run-off triangle. A notable exception
is  \citeA{mack93}
{\color{black}  as well as  Bayesian adaptations such as  \citeA{taylor15}, \citeA{peters2017full}, and \citeA{boratynska22}; 
and extensions to multiple triangles, e.g.,  \citeA{taylor15}, \citeA{schnieper91} and \citeA{wuthrich18}. }
However, as has been noted in \citeA{bischofberger2020continuous}
and \citeA[Chapter 11.2]{Mikosch:09}
the assumptions of the model in \citeA{mack93} do not conform well with the data-generating mechanism from an individual claim level perspective:
claim payments in later development periods are assumed to always stem from earlier payments. 


{\color{black} While reserving in the insurance practice is traditionally modeled using run-off triangles, there is a research stream that started with the seminal works of \citeA{arjas89}, \citeA{norberg93}, \citeA{haastrup96} and \citeA{antonio14} that uses the individual data from which development triangles stem.  Example of recent works in this area include \citeA{denuit17}, \citeA{delong22}, \citeA{fung22}, \citeA{robben22}, \citeA{crevecoeur23}, \citeA{michaelides23}, \citeA{bucher23}, \citeA{yanez2024modeling}, and \citeA{yang24}.}

In recent years there have been several papers that model continuous run-off triangles based on individual claims data. In \citeA{miranda13} and \citeA{hiabu2016sample}  the loss-triangle based reserving-problem is translated into a continuous framework and the authors propose to estimate a density function corresponding to claim  counts via kernel smoothers. The model has been extended in 
\citeA{lee15}, \citeA{lee17}, and \citeA{mammen21} to account for 
seasonal effects, operational time and calendar effects, respectively. 
\citeA{hiabu17} and \citeA{bischofberger2020continuous} model the claim development and show an asymptotic relationship between chain-ladder's development factors and a time-reversed hazard rate.
However, in contrast to our paper, the estimation in \citeA{hiabu17} and \citeA{bischofberger2020continuous} is based on individual claims data and the relationship between chain-ladder's development factors and a time-reversed hazard rate is based on continuous observations only. In summary the contribution of our paper are as follows
\begin{itemize}
\item We propose an estimator of the claim development based on data given as a run-off triangles. While pre-smoothing into run-off triangles is theoretically less efficient than direct estimation based on individual claims data, our proposed estimator can be applied on data readily available and most familiar to reserving actuaries.


\item  We establish that modeling the claim development via an age-model replicates chain-ladder estimates. This is achieved by deriving an exact and non-asymptotic relationship between time-reversed  averaged hazard rates and chain-ladder's development factors. 

\item We show that the link between time-reversed hazard rates and chain-ladder's development factors is akin to the link between central mortality rates and survival probabilities in the context of mortality modeling.  Thus, in contrast to \cite{tsai22} who leverage claim reserving methods to project mortality rates, we leverage age-period-cohort mortality modeling methods to  estimate claim reserves. 

\item By using discrete estimators, we show how existing GLM-based estimators within the age-period-cohort framework 
can be employed to model claim development. Indeed, this paper comes with the \texttt{clmplus} package  \cite{clmplus}, an out-of-the-box set of tools available in R to practitioners and researchers that may want to apply or build on our framework. The \texttt{clmplus} implementation relies on the tools available in the \texttt{StMoMo} package \cite{stmomo}, that made the the age-period-cohort framework easy and ready to use in a mortality context.

\item In the case studies on real datasets presented in Sections \ref{sec:ModelComparison} and \ref{sec:naic}, we find  that our proposal of modeling the claim development via an age-period-cohort  models  seems to outperform age-period-cohort models based on claim amounts in most cases. However, there are exceptions and we are not claiming that our framework should replace other modeling techniques but rather complement them.


\end{itemize}

The rest of the paper is organized as follows. In Section \ref{sec:prelminaries} we discuss some preliminary ideas of our modeling approach, emphasizing its connections with the mortality modeling literature and making a parallel with classical claim amount based GLM approaches replicating the chain-ladder.  
In Section \ref{sec:ClaimDeveloment}, we introduce the general modeling framework underpinning our proposed approach, show how to define a model that replicates the chain-ladder estimates, and discuss additional effects that can be added in the modeling. In Section \ref{sec:da}, we present our first data application and show how to use the tools available in \texttt{clmplus} to choose the best model to compute the claims reserve. 
 To further validate our results, in Sections \ref{sec:ModelComparison} and \ref{sec:naic}, we compare our models to the age-period-cohort models  within the classical GLM framework targeting claim amounts \cite{harnau18}. 

\section{Preliminaries, connections to mortality modeling and summary of main results}\label{sec:prelminaries}

We propose to model the claim development via a GLM  which -- as we will see -- in its simplest form can replicate chain-ladder's development factors but additionally allows for straightforward variations by changing the underlying distribution or structure.
To this end,  we consider
a run-off triangle 
$$\mathcal{X}=\{X_{kj}: k,j=0, \ldots, m ; k+j \leq m\}, \quad m>0,$$
where $X_{kj}$ denotes the observation for accident period $k$ and development period $j$. {\color{black} The calendar period is then $k+j$ and $m$ is the evaluation date}.
We do not further specify  the nature of the triangle. Two possibilities are
$\mathcal X$ being an aggregation of incremental claim counts or claim payments. In the first case the development period $j$ denotes the delay from accident to report and in the second case  it denotes the delay from accident to payment.
We propose to consider the observed claim development, {\color{black} $\widetilde \mu_{kj}$}, for $j=1, \ldots, m$,

{\color{black} 



\begin{align}\label{muhat}
\widetilde \mu_{kj}=\frac{X_{kj}} {\sum_{l<j} X_{kl}+ \frac 1 2 X_{kj}}=\frac{X_{kj}}{E_{kj}}, \quad E_{kj}=\sum_{l<j} X_{kl}+ \frac 1 2 X_{kj},
\end{align}

}
within the generalized linear models (GLM) framework.  
If $\mathcal{X}$ are claim counts, 
then {\color{black} $\widetilde \mu_{kj}$}, can be motivated as the estimator of  an averaged hazard rate that  runs in reversed development time, denoted $\mu_{kj}$. This can be seen by noting that {\color{black} $\widetilde \mu_{kj}$} is the ratio of occurrence and expected exposure. We remark that in this manuscript we use the term \textit{exposure} to denote the quantity $E_{kj}$ defined in Equation \ref{muhat}.
In the case of $\mathcal{X}$ being claim payments, $\mu_{kj}$ is additionally  weighted by the expected claim size in cell $(k,j)$. An explicit and more technical definition of $\mu_{kj}$ in the latter case will be provided in the next section. A simple rearrangement shows that for $k=0,\dots, m; j=1,\dots,m$, the observed claim development, {\color{black}  $\widetilde \mu_{kj}$}, has the following relationship to the  individual development factors, {\color{black} $
\widetilde f_{kj}:={\sum_{l\leq j}X_{kl}} / {\sum_{l<j} X_{kl}}:$}

{\color{black}


\begin{align}\label{transf}
\widetilde f_{kj}= \frac{2+\widetilde \mu_{kj}}{2-\widetilde \mu_{kj}}.
\end{align}
}

Equation  \eqref{transf} should be seen as general formula to switch between a modeled claim development and the associated development factor; with the latter being subsequently used for prediction. More concretely, starting from the raw observations {\color{black} $\tilde \mu_{kj}$}, we will assume some structure on $\mu_{kj}$ and derive estimates {\color{black} $\widehat \mu_{kj}$} of the claim development.
Thereafter one applies the equivalent of   \eqref{transf} to derive estimates of the development factors, {\color{black} $\widehat f_{kj}$}, which are subsequently used to project observations  $\mathcal X$
into the lower triangle.

Interestingly, there is a parallel to draw to mortality modeling in actuarial science and demography. {\color{black} When modeling mortality, we usually consider the force of mortality and the central mortality rate. The force of mortality is the instantaneous mortality hazard rate at a given age, while the central mortality rate can be thought of as the average force of mortality (average hazard rate) over a one-year age interval. Formally, let $m_{kj}$ denote the central mortality rate at age $j$  for cohort $k$, that is, the average force of mortality between age  $j$ and $j+1$ for cohort $k$. Also, let $q_{kj}$ denote the probability that an individual aged $j$ in cohort $k$ dies before reaching age $j+1$, and $p_{kj}=1-q_{kj}$ the probability that an individual in cohort $k$ who is age $j$ survives to age $j+1$.}
The central mortality rate $m_{kj}$ is not observed and estimation is based on the observed rates {\color{black} $\widetilde m_{kj}= d_{kj}/e_{kj}$} , where $d_{kj}$ denotes the observed number of deaths and $e_{kj}$ denotes the person-years at risk. The quantity $e_{kj}$ is often approximated as the number of people of cohort $k$ that died at age $j+1$ or later plus one half times the number of people who died at age $j$. Notice that formula \ref{muhat} uses the same approximation.

In mortality modeling, one {\color{black} common approach is to model the} central mortality rate $m_{kj}$ using an age-period-cohort framework leading to {\color{black} model} estimates {\color{black}  $\widehat m_{kj}$}. Afterwards, {\color{black} to obtain one year death probabilities and survival probabilities}, one calculates the discrete quantities {\color{black} $\widehat q_{kj}$ and $\widehat p_{kj}$} using plug-in estimates in the following standard life-table relationships \cite{chiang72}:
\begin{align}
\widehat q_{kj} &= \frac{2\widehat m_{kj}}{2+\widehat m_{k,j}}\label{mxtoqx} \\
\widehat p_{kj} &= 1 - \widehat q_{kj} = 1- \frac{2\widehat m_{kj}}{2+\widehat m_{kj}} = \frac{2-\widehat m_{kj}}{2+\widehat m_{kj}} \label{mxtopx}.
\end{align}
Comparing Equation \eqref{mxtopx} and  Equation \eqref{transf} we can summarize the following: 
\begin{itemize}
    \item In mortality modeling we estimate the central mortality rate $m_{kj}$ from  observations
{\color{black} $\widetilde m_{kj}$} leading to estimates {\color{black} $\widehat m_{kj}$}
and use them to derive estimates for  one year conditional survival probabilities {\color{black}$\widehat p_{kj}$}.

 \item For claims reserving we propose to estimate the claim development $\mu_{kj}$ from observations {\color{black}$\widetilde \mu_{kj}$} leading to estimates {\color{black}$\widehat \mu_{kj}$} and use it to calculate  development factors {$\widehat f_{kj}$} which are estimates of inverse conditional probabilities. 
\end{itemize}
{\color{black} Table \ref{tab:parallel_mortality_claims} further summarizes the parallel between mortality modeling and claims reserving.  We will exploit this parallel to provide more flexible claim development models and leverage existing mortality modeling software.}

\begin{table}[ht!]
\centering
\renewcommand{\arraystretch}{1.3} 
\begin{tabular}{|c|l|p{6cm}|}
\hline
\textbf{Concept} & \textbf{Mortality Modeling} & \textbf{Claims Reserving} \\ \hline

\multirow{3}{*}{\textbf{Rate}} 
    & $m_{kj}$, $\widetilde m_{kj}$, $\widehat m_{kj}$ 
    & $\mu_{kj}$, $\widetilde \mu_{kj}$, $\widehat \mu_{kj}$ \\ \cline{2-3}
    & Central mortality rate & Claims development rate \\ \cline{2-3}
    & Average force of mortality 
    & Averaged hazard rate in reversed development time \\ \hline

\multirow{3}{*}{\textbf{Probability}} 
    & $p_{kj}$, $\widetilde p_{kj}$, $\widehat p_{kj}$ 
    & $f_{kj}$, $\widetilde f_{kj}$, $\widehat f_{kj}$ \\ \cline{2-3}
    & One-year survival probability & Claim development factor \\ \cline{2-3}
    & Conditional survival probability 
    & Inverse conditional probability \\ \hline

\end{tabular}
\caption{\color{black} Parallel between mortality modeling and claims reserving quantities.}
\label{tab:parallel_mortality_claims}
\end{table}




As such, we will borrow in the sequel the language from mortality modeling and Lexis diagrams \cite{carstensen2007age}, with $k$ denoting cohort, $j$ age and $k+j$ period. In claims reserving, age $j$ usually corresponds to development period, period $k+j$ to calendar time and cohort $j$ to accident period, see Table \ref{tab:mortalitynonlife}.

\begin{table}[H]
\centering
\caption{Notation parallel, Lexis dimensions and claims reserving.} 
\begin{tabular}{c r r}
Notation & Claims reserving & Lexis dimensions \\
\hline
$j$         & development period & age \\
$k$      & occurrence or accident date & cohort \\
$k+j$     & calendar date& period \\
\hline
\end{tabular}
\label{tab:mortalitynonlife}
\end{table}


We note now that if we model the underlying development,  $\mu_{kj}$, by an age-model, i.e., $\mu_{kj}=a_j$, and assume that $X_{kj}$ given $E_{kj}$ follows an (over-dispersed) Poisson distribution, we can exactly replicate the chain-ladder point estimates. 
The replication is achieved by applying formula \eqref{transf} to move from the modeled claim development, {\color{black}$\widehat \mu_{kj}=\widehat a_j$}, to the development factors: {\color{black} $\widehat f_j= (2+\widehat a_j)/ (2- \widehat a_j)$}. The resulting {\color{black} $\widehat f_j$} equals  exactly chain-ladder's development factors.
In Table \ref{tab:overview}  below, we contrast our approach to the classical GLM approach in claims reserving where the claim amount is modeled.
\begin{table}[H]
\centering
\caption{GLM Models replicating chain-ladder estimates} 
\begin{tabular}{r|rr}
&Our proposal & Classical GLM approach\\
\hline
Modeling of &  claim development &  claim amount\\
Structural assumption &$\mu_{kj}=a_j$ &$\ E[X_{kj}]= \eta_k\nu_j$\\
Distributional assumption &$X_{kj}|E_{kj} \sim \textrm{Pois}(E_{kj}a_j)$ &$X_{kj} \sim \textrm{Pois}(\eta_k\nu_j)$\\
&$E_{kj}=\sum_{l<j} X_{kl}+ \frac 1 2 X_{kj}$\\
\hline
\end{tabular}
\label{tab:overview}
\end{table}

Noticeably, the age-model within our proposed modeling framework only uses half the number of parameters compared to the classical GLM approach. That is because we do not model exposure, but only the claim development and then have to estimate only $a_j$ rather than $\eta_k$ and $\nu_j$; see also \cite{mack2000comparison, verrall2000comments} for a related discussion between Mack's model and  GLM modeling on claim amounts.
One advantage of our simpler model is that more flexible models can be easily incorporated.
In the accompanying \texttt{clmplus} R-package \cite{clmplus}, one can fit the claim development to any model from the age-period-cohort family commonly used in the mortality modeling context  \cite{hunt21}. We hope that the new modeling structure and the additional flexibility does not only improve prediction  but that it also provides a framework for actuaries to better analyze and compare model fit. 
\newline

\section{Modeling the claim development}\label{sec:ClaimDeveloment}

 Having established the main results of our proposed modeling approach, we provide in this section further theoretical details underpinning our approach.  In the sequel, we will consider for $k=0,\dots,m;$ $j=1,\ldots,m$, $\eta \in (0,1)$ the following quantity

{\color{black}

\begin{equation}
\label{eq:mukj}
\widetilde \mu_{kj}(\eta)=\frac{X_{kj}}{E_{kj}(\eta)},
\end{equation}

}

where
\[
 E_{kj}(\eta)=\sum_{l<j} X_{kl}+ \eta X_{kj}.
\]
Note that in the previous section we only considered the special case $\eta=0.5$, but here we make explicit that {\color{black} $\widetilde \mu_{kj}(\eta)$} depends on $\eta$. The factor $\eta$ accounts for how much claims in $(k,j)$ contribute to the exposure in $(k,j)$; it is also known as lost exposure in mortality modeling \cite[p.~66]{wilmoth07}.
The value of $\eta$ cannot be estimated without additional information and $\eta=0.5$
corresponds to assuming a uniform distribution of claims within a cell. 
Observe that we are not modeling {\color{black} $\widetilde \mu_{kj}(\eta)$} for $j=0$, because a run-off triangle (i.e. aggregated data) does not provide any information on the exposure in the first column: $E_{k0}(\eta)$ as fraction of $X_{kj}$ is completely determined by $\eta$.
To illustrate the rational of modeling {\color{black}$\widetilde \mu_{kj}(\eta)$}, let's
assume $\mathcal X$ being claim payments and  $X_{kj}$ stemming from iid payments $(Z_i,U_i,T_i)$, $i=1,\dots,n$, where 
$Z_i$ is the payment size, $U_i$ accident date and $T_i$ the delay between accident and payment. In Appendix \ref{appendix:le}, we discuss that 
{\color{black} $\widetilde \mu_{kj}(\eta)$} can be motivated as estimator of $\mu_{kj}$ -- an exposure weighted average of the expected payment size weighted hazard rate in reversed development time $\alpha^{\ast,R}(t|u)$:

\begin{align*}
\alpha^{R}(t|u)&= \lim_{h \downarrow 0} h^{-1}{\mathbb P}\left(T_i\in(t-h,t]|\ T_i \leq t < \mathcal T-U_i , U_i=u\right), \\
\alpha^{\ast,R}(t|u)&=
\frac{  \mathbb E[Z_1 | T_1=t, U_1=u]}{ \mathbb E\left[Z_1 | \ T_1 \leq t, U_1=u\right]}\alpha^{R}(t|u),\\
\mu_{kj}&=\frac{\delta \int_{\mathcal{P}_{kj}} \alpha^{\ast,R}(s|u)p_U(u)\gamma(s,u) ds du} {\int_{\mathcal{P}_{kj}} p_U(u)\gamma(s,u) ds},
\end{align*}

where $\gamma(s,u)=\mathbb E[Z_iI(T_i \leq s < \mathcal T-U_i),
| U_i=u]$ and  $p_U$ is the marginal density of $U_i$. The 
average runs  over the parallelogram $\mathcal P_{kj}=\{(t,u): t_j+u_k-u\leq t \leq t_{j+1}+u_k-u; u\in [u_k,u_{k+1} ), t\geq 0\}$ with  an equi-distant grid $t_0=0,\dots, t_{m+1}=\mathcal T$ and $u_0=0,\dots, u_{m+1}=\mathcal T$, where $t_j-t_{j-1}=u_k-u_{k-1}=\delta; \delta>0; j,k=0,\dots,m$. 
\newline
\noindent 

\begin{remark}
One may ask why we propose to model $\mu_{kj}$ and not the individual development factors $f_{kj}$ or the actual hazard rate $\alpha^{\ast,R}(t|u)$. We discuss both alternatives consecutively:

\begin{itemize}
    \item  Why not modeling $f_{kj}$?  The individual development factors stem from a discrete world and are not defined in a continuous setting. The claim development $\mu_{kj}$ however is up to a constant $\delta$ well defined in a continuous setting in the sense that $\mu_{kj}/\delta$  converges to $\alpha^{\ast,R}(t|u)$ for the grid-size $\delta$ converging to zero. This is desirable from an abstract point of view but also from a modeling perspective  because it enables to embed the claim development in a wider continuous modeling framework.
 \item Why not model the continuous hazard rate   $\alpha^{\ast,R}(t|u)$ directly? This is because we start from run-off triangles. While one could still impose a structure on $\alpha^{\ast,R}(t|u)$ first, the implications for $\mu_{kj}$ are not easy to derive and may complicate things unnecessarily. As mentioned in the previous section, there is an interesting parallel to draw to the field of mortality modeling in actuarial science and demography where usually an averaged version of the hazard rate (force of mortality) is being modeled.  The force of mortality is the central object for further modeling and interpretation and it is also used to estimate the conditional probability of dying, see e.g. the methods protocol for the Human Mortality Database \cite{wilmoth07}.
 In a similarly fashion, we propose to estimate the development factors after modeling $\mu_{kj}$.
\end{itemize}

\end{remark}

In the previous section we discussed that for $\eta=0.5$, there is a one to one relationship between chain-ladder's individual development factors, {\color{black} $\widetilde f_{jk}$}, and {\color{black} $\widetilde \mu_{kj}(\eta)$}. An analogue relationship remains true for general $\eta$. For $k=0,\dots; j=1,\dots,m$,
the individual development factors are defined as
{\color{black} $
\widetilde f_{kj}={\sum_{l\leq j}X_{kj}} / {\sum_{l<j} X_{kl}},
$}
hence straight forward calculations (see Appendix \ref{appendix:fkjmukj}) lead to

{\color{black}
\[
\widetilde f_{kj}= \frac{1+(1-\eta) \widetilde \mu_{kj}(\eta)}{1- \eta \widetilde\mu_{kj}(\eta)}.
\]
}
Note that by definition the individual development factors do not depend on $\eta$, i.e.,
for two values $\eta, \eta', \eta\neq \eta'$,
{\color{black} $
\widetilde f_{kj}=
\{1+(1-\eta) \widetilde \mu_{kj}(\eta)\}/\{1- \eta \widetilde\mu_{kj}(\eta)\} =  \{1+(1-\eta') \widetilde \mu_{kj}(\eta')\}/\{1- \eta' \widetilde\mu_{kj}(\eta')\}={\sum_{l\leq j}X_{kj}} / {\sum_{l<j} X_{kl}}.
$}
However this does not mean that predictions, as will be defined in Section \ref{sec:pred}, are not affected by $\eta$. 
The reason for this is that predictions will be based on some predicted development factors
{\color{black} $\widehat f_{kj}$} rather than on {\color{black}$\widetilde f_{kj}$} which will be calculated from
some predicted claim development {\color{black} $\widehat \mu_{kj}$} rather than  {\color{black} $\widetilde \mu_{kj}(\eta)$}.
The invariance of {\color{black}$\widetilde f_{kj}$} to different choices of $\eta$ gives hope that {\color{black} $\widehat f_{kj}$} is not affected much by different $\eta$. Indeed in the cases we investigated, we found that changing the value of $\eta$ only lead to minor changes in the predicted development factors and overall predictions. Therefore, we conjecture that the actual choice of $\eta$ is {\color{black}often} of minor importance. {\color{black}Nevertheless, in a practical application one may choose a specific $\eta$ based on model validation and 
expert's knowledge. In the latter case, for example if more claims are expected to occur in the first half of a period than in the second half of a period, this would indicate that $\eta>0.5$.}




\subsection{A stochastic model replicating chain-ladder}
\label{ss:stochasticcl}

Let us assume that for $k=0,\dots,m$ and $j=1, \ldots, m$,

\begin{equation}
\label{eq:agemodel}
 \mu_{kj}= a_j. \qquad [\text{age-model}]    
\end{equation}

\noindent Furthermore, we assume that  
the entries $X_{kj}$ are independent given $E_{kj}$ and  for $k=0,\dots,m; j=1,\dots, m,$ they  follow a Poisson distribution:
\[
X_{kj}|E_{kj} \sim \text{Pois}(O_{kj}), \quad O_{kj}:= E_{kj} \mu_{kj}.
\]
\begin{remark}
Analogue to what is done in the classical GLM reserving literature,
one can change the Poisson assumption to an overdispersed Poisson assumption without altering the point estimates, see \citeA[p.~323]{mccullagh19}.
\end{remark}



This model, which we will refer to as the age-model, assumes that the claim development depends only on age, that is, on the development period of the claim (recall Table \ref{tab:mortalitynonlife}). Then the log-likelihood is given by

\begin{align*}
    l\left(a_1,\dots a_m \mid X_{kj}, E_{kj} , j=1,\dots, m; j+k\leq m\right) \propto \sum_{j, k}  X_{kj} \log( a_j E_{kj}) - E_{kj}a_j,
\end{align*}

leading to the first order condition

\[
 \sum_{ k} \frac{ X_{kj}}{ \widehat a_{j}} - E_{kj}=0,
\]

with minimizer  

{\color{black}


\begin{align*}
 \widehat \mu_{kj}= \widehat a_j =\frac{\sum_{k} X_{kj}}{\sum_{k} E_{kj}}.
\end{align*}

}

In particular, for $j=1,\dots,m$,

{\color{black}


\begin{align}
\label{eq:aj2fj}
\widehat f_j:=\frac{1+ (1-\eta)\widehat a_j}{1- \eta \widehat a_j}=\frac{ \sum_{l\leq j}\sum_{k=0}^{m-j} X_{k l}}{ \sum_{l< j}\sum_{k=0}^{m-j} X_{k l}}=f_j^{\text{chain-ladder}},
\end{align}

}

showing how the age-model replicates chain-ladder's development factors for any choice of $\eta$. The complete proof can be found in Appendix \ref{appendix:agecl}. 
 Hence, the age-model is a special case where the predicted development factors do not depend on 
 $\eta$.
From now on, we will assume that claims in the parallelograms occur uniformly, using $\eta = \frac 1 2$.

\begin{remark}
Note that minimizing the Poisson likelihood leads to the same minimizer as minimizing the weighted least squares
\[
 \sum_{j,k} E_{kj }(\hat \mu_{kj}- \hat a_j)^2.
\]
\end{remark}

\subsection{Further models}

In the previous  subsections, and in contrast to existing literature, we modeled the claim development and not the claim amount. 
A generalized framework for claim amounts via an age-period-cohort construction has been considered in  \citeA{kuang08b, kuang08}, \citeA{harnau18}, and \citeA{nielsen20}. 
Apart from its simplicity, one further advantage of our claim development formulation is that we can exploit the connection with mortality rate modeling and extend Equation \eqref{eq:agemodel} to any model from the family of age-period-cohort (APC) stochastic hazard rate models.
We now assume that the claim development has the following form:


\begin{equation}
\label{eq:gapc}
    \log(\mu_{kj})=a_j+  c_{k+j} + g_{k}.
\end{equation}

\noindent 

Note that by some misuse of notation,  $a_j$ now denotes the age component on the log-scale while in the previous section it was the age component on the original scale.  The logarithmic link function, connects the estimator for the hazard $\mu_{kj}$ (our response variable) to the effects specified in the linear equation on the right-hand side. This equation defines a APC model for the claims development. By doing so, it captures the following components:

\begin{itemize}
    \item $a_j$ is the age effect (development period) on $\mu_{kj}$.    
    \item $c_{k+j}$ is the period effect (calendar date) effect on $\mu_{kj}$.
    \item $g_k$ is the cohort effect  (accident date)  on $\mu_{kj}$.
\end{itemize}

The theory for age-period-cohort models given our triangular structure is well understood \cite{kuang08, kuang08b}.

{\color{black} Each row of Table \ref{tab:presetmodels} describes an APC model. The first column contains the model label from the \texttt{clmplus} package and the second column contains the model description. The third column shows the effects of the APC model. }
\newline

APC models are identifiable up to an identification constraint. We show in column four of \autoref{tab:presetmodels} the default identification constraints that we used in this paper and in our \texttt{clmplus} package. The age-model is identifiable without additional constraints on the (categorical) effect $a_j$ as the linear predictor of the Poisson regression model in \autoref{eq:gapc} contains no intercept. For the age-cohort model, we set $g_0=0$. For the age-period model, we set $c_1=0$. 
The age-period-cohort model is invariant with respect to the following two parameter transformations:

\begin{align*}
    \left(a_j, c_{k+j}, g_k\right) &\rightarrow\left(a_j+\phi_1-j\phi_2 , c_{k+j}+(k+j)\phi_2, g_k-\phi_1-k\phi_2 \right),
    \\
    \left(a_j, c_{k+j}, g_k\right) &\rightarrow\left(a_j+\psi_1, c_{k+j}-\psi_1, g_k\right), 
\end{align*}

with $\phi_1, \phi_2, \psi_1 \in \mathbb{R}$ \cite{kuang08}. {\color{black} To ensure identifiability of the age-period-cohort model, we follow the usual practice in mortality modeling (see, e.g., \citeA{haberman11}) and impose} 
\begin{align*}
 \sum^{m}_{s=1} c_{s}=0, \quad \sum^{m-1}_{k=0} g_k=0, \quad \sum^{m-1}_{k=0} k g_k=0.
\end{align*}
The constraints on the cohort effect ($\sum_{k} g_k=0, \sum_{k} k g_k=0$) imply that $g_k$, for $k=0,\dots, m-1$,  fluctuates around zero without any  trend.

\begin{table}[H]
\caption{\label{tab:presetmodels} In the first column we cite the model label. The second columns provides a description of the Lexis diagram dimensions: age is development period, cohort is accident period and period is calendar time. The effects are displayed in the third column. In the last column we add the identification constraints on the effects displayed in column three. }
\centering
\begin{adjustbox}{width=15cm,center}
\begin{tabular}{rrrr}
\hline
\texttt{clmplus} (short) &Lexis dimensions& Effects& default identification constraints\\
\hline
 a&age (chain-ladder model)&$a_j$&-\\
ac&age--cohort&$a_j+g_{k}$&$g_{0}=0$\\
ap&age--period&$a_j+c_{k+j}$&$c_{1}=0$\\
apc&age--period--cohort&$a_j+c_{k+j}+g_{k}$&  $\sum^{m}_{s=1} c_{s}=\sum^{m-1}_{k=0} g_k=\sum^{m-1}_{k=0} k g_k=0$\\
\hline
\end{tabular}
\end{adjustbox}
\end{table}

{\color{black} \begin{remark}
Modeling a calendar time effect and/or an accident time effect on the claims development can help capture environmental changes that traditional approaches such as the chain ladder may not capture. By environmental changes, we mean changes in the insurer's underwriting process or external events that might affect claims in the same accident period or calendar period. In the actuarial literature, this issue has gained new attention in the context of the COVID pandemic and has been discussed in detail in the works of \citeA{okine22, okine23} and \citeA{riegel23} and in the data application of \citeA{almudafer21}. In Section \ref{ss:environment2} and Appendix \ref{appendix:simulationstudy}, we provide empirical evidence that our model can outperform standard reserving models on the same data used in \citeA{almudafer21}.
\end{remark}}

\subsection{Extrapolation of cohort and period effects}
\label{ss:effectsextr}

As discussed in \citeA{kuang08b}, it is necessary to have a linear trend or a random walk with a drift to have identification invariant forecasts on the lower triangle.
The dynamics of the predicted development factors when using hazard models with a cohort effect will depend on $g_k$ with $k=0, \ldots, m$. Having no information about the exposure in $j=0$, we only have 
predictions $\hat g_1, \dots \hat g_{m-1}$.
Hence, one needs to extrapolate the cohort effect  for the last row, $\widehat g_m$. In the case studies that we will present, we will show the results from the models in Table \ref{tab:presetmodels}.  We will assume that the cohort component follows an ARIMA (1,1,0) model with drift $\nu_0$:

\begin{equation}\label{ex1}
    g_k   =  \nu_0  + g_{k-1}+  \phi (g_{k-1} - g_{k-2}) + \xi_{k}, \qquad \xi_{k} \sim N(0, \sigma),
\end{equation}

with $\phi \in \mathbb{R}$.
For the models in Table \ref{tab:presetmodels} that include a period component we will need to extrapolate the period effect for the future calendar years, $\widehat c_{m+s}, s = 1, \ldots, m-1$. In the next sections we will model the period effects as a random walk with drift $\nu_1$:

\begin{equation}\label{ex2}
    c_{k+j}=\nu_1 + c_{k+j-1}+\xi_{k+j} \qquad \xi_{k+j}, \sim N(0, \sigma).
\end{equation}

Note that while here  we use \eqref{ex1} and \eqref{ex2} to extrapolate cohort and period effects, in practice the models for extrapolating cohort and period effects should be a data-driven; possibly in conjunction with expert judgement. 


\subsection{Prediction} \label{sec:pred}

Based on Equation \eqref{eq:gapc},  fitted values are given by

{


\begin{align*}
    \widehat \mu_{kj}= \operatorname{exp}\left( \widehat a_j+ \widehat c_{k+j}+ \widehat g_{k}\right).
\end{align*}
}
Thus, for the upper triangle,  $0\leq (k,j)$ with $k+j\leq m$, we define the fit as

{\color{black}

\begin{align*}
    \widehat X_{kj}= E_{kj}\widehat \mu_{kj}.
\end{align*}
}

If one wishes to get predictions for the the lower triangle, $k+j>m$, an obstacle is that the exposure $(E_{kj})$ is not observed there.
A solution is to use the chain principle known from the chain-ladder method.
To this end, define the predicted development factors for accident year $k=0,\dots,m$ and development year $j=1,\dots,m$ as,
{\color{black}



\begin{align}
\label{eq:fittedfkj}
    \widehat f_{kj}= \frac{(2+\widehat \mu_{kj})}{(2-\widehat \mu_{kj})}.
\end{align}

}
Note that this is analogue to what is done in mortality prediction when the fit for the force of mortality is transformed to conditional survival or death probabilities (cf. Equations \ref{mxtoqx} and \ref{mxtoqx}).
Next, define the cumulative data as

\begin{align*}
    C_{kj}= \sum_{s\leq j} X_{k s}.
\end{align*}






An estimate for the lower triangle $(k+j>m)$ is derived via
\begin{align*}
    \widehat C_{kj}= C_{k,m-k}\prod^j_{l=m-k+1}  \hat f_{kl}, \qquad \widehat X_{kj}= \begin{cases}
    \widehat C_{kj}-  C_{k,j-1} & \text{if}\  k+j=m+1,\\
    \widehat C_{kj}-\widehat C_{k,j-1} & \text{if} \ k+j\geq m+2.
    \end{cases}
\end{align*}


\section{Empirical study}
\label{sec:da}

In Section \ref{ss:autobisuper} we illustrate our framework using the \texttt{AutoBI} dataset available in the \texttt{R} package \texttt{ChainLadder} \cite{ChainLadder} . For completeness, we include this dataset in Appendix \ref{appendix:autobi}.
{\color{black} In Section \ref{ss:environment2}, we present a simulation study that illustrates our models behavior when data are affected by environmental changes. Further details on the data simulation are presented in Appendix \ref{appendix:simulationstudy}. In Section \ref{sec:ModelComparison}, we present a model selection strategy and show an application on the $30$ publicly available datasets listed in Appendix \ref{appendix:triangleslist}. The model selection strategy is applied to historically fully developed market data in Section \ref{sec:naic}. In  Appendix \ref{appendix:clmpluscode} shows how to use the \texttt{clmplus} package to replicate some of the results in this section.
}

\subsection{Illustrations about our model on the the AutoBI data set }
\label{ss:autobisuper}

Using the \texttt{AutoBI} dataset, we first show that both the GLM approach based on claim amounts \cite{england99} and the GLM based on claim developments can replicate chain-ladder. In the latter case, as discussed earlier, this is achieved by specifying the claim development $\mu_{kj}$ with the age-model in Equation \eqref{eq:agemodel} and exploiting the relation in Equation \eqref{eq:aj2fj} together with the chain-ladder algorithm to forecast the claims development.
After that, we will show the improvement that can be gained by adding a cohort component.
Lastly, a discussion on models that require a period component will follow.  
In this second part, the different claim development fits are compared in terms of scaled deviance residuals on the development. The scaled deviance residuals $r_{kj}$ for cohort $k$ and age $j$ are:
\begin{align*}
    r_{kj}=\operatorname{sign}\left(X_{kj}-\widehat{X}_{kj} \right) \sqrt{\frac{\operatorname{dev}(k, j) (K-\nu)}{D}}
\end{align*}

\noindent where:

\begin{align*}
    \operatorname{dev}(k, j)=2\left[X_{kj} \log \left(\frac{X_{kj}}{\widehat{X}_{kj}}\right)-\left(X_{kj}-\widehat{X}_{kj}\right)\right],
\end{align*}

\noindent $D=\sum_k \sum_j \operatorname{dev}(k, j)$, $K$ is the number of observations, and $\nu$ the number of parameters in the model. The ratio ${D}/(K-\nu)$ estimates the dispersion parameter. By providing a heat-map of the scaled deviance residuals $r_{k,j}$ on the upper triangle any undesirable pattern can be detected and removed.

\subsubsection{Replicating the chain-ladder via an age-model}
\label{ss:clm}


Table \ref{tab:estimates} shows the results of applying three methods to estimate the claims reserve in the \texttt{AutoBI} dataset, namely, the chain-ladder reserve as implemented in the \texttt{ChainLadder package}, the age-cohort GLM approach based on claim amounts implemented in the \texttt{apc} package \cite{package:apc}, and the age-model \eqref{eq:agemodel} within our proposed framework that models the claim development as implemented in our \texttt{clmplus} package. Here, we see that all three approaches yield the same  reserve estimates in each cohort. 

\begin{table}[H]
\caption{\label{tab:estimates} Comparison between the reserve from the chain-ladder method, the GLM approach based on claim amounts and  our proposal, \texttt{clmplus}, that models the claim development via a GLM on the \texttt{AutoBI} data.
}
\centering
\begin{tabular}{rrrr}
\hline
accident year & chain-ladder & GLM(claim amounts) & \texttt{clmplus}\\ 
 & - & age-cohort-model &age-model \\ 
  \hline
  0 & 0.00 & 0.00 & 0.00 \\ 
  1 & 67.24 & 67.24 & 67.24 \\ 
  2 & 345.19 & 345.19 & 345.19 \\ 
  3 & 940.69 & 940.69 & 940.69 \\ 
  4 & 2350.86 & 2350.86 & 2350.86 \\ 
  5 & 4466.77 & 4466.77 & 4466.77 \\ 
  6 & 9103.24 & 9103.24 & 9103.24 \\ 
  7 & 14480.44 & 14480.44 & 14480.44 \\ 
   \hline
\end{tabular}
\end{table}

\begin{figure}[H]
    \centering
    \caption{\label{fig:paramsestimates} Results from the different models fitted on the \texttt{AutoBI} dataset. From left to right: the fit of the $a_j$ effect in Equation \ref{eq:agemodel}, the chain-ladder development factors in Equation \ref{eq:aj2fj}, as well as the age effect and the cohort effect when modeling the claim amount. It follows from the previous sections that the results in \ref{fig:clm} can be obtain from those in \ref{fig:clmplus} with the transformation in Equation \eqref{eq:aj2fj}. }
    \begin{subfigure}[t]{0.49\textwidth}
        \centering\includegraphics[width=\linewidth]{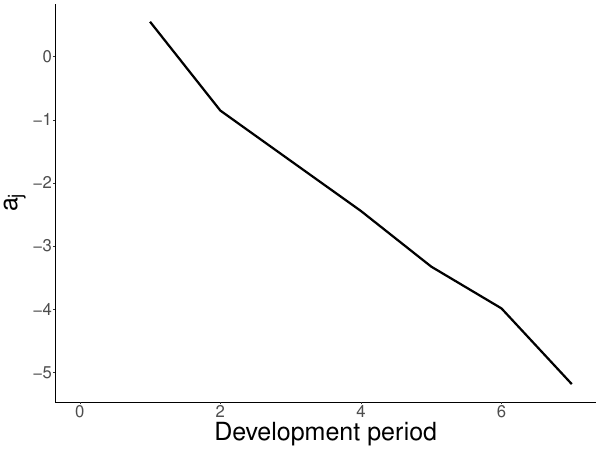}
        \caption{\label{fig:clmplus} Age effect plot from the age-model (our proposals).}

    \end{subfigure}
    \hfill
    \begin{subfigure}[t]{0.49\linewidth}
        \centering\includegraphics[width=\linewidth]{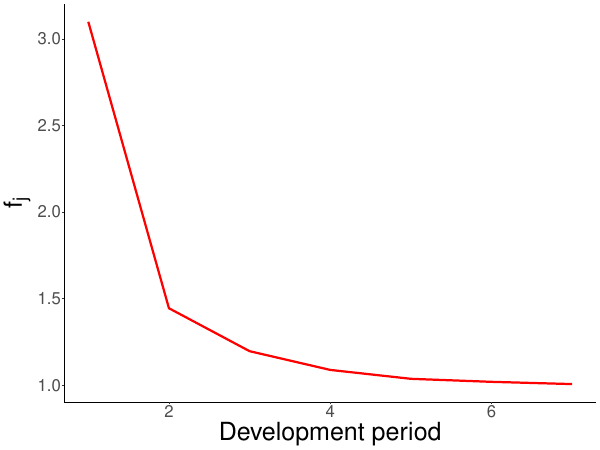}
        \caption{Development factors estimated from the chain-ladder method.}
        \label{fig:clm}
    \end{subfigure}
    \hfill
    \begin{subfigure}[t]{0.49\linewidth}
        \centering\includegraphics[width=\linewidth]{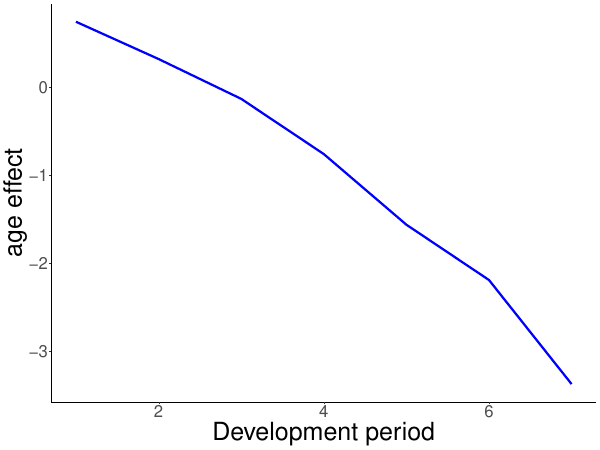}
        \caption{Age effect estimated from the GLM approach modeling the claim amount in a cross-classified ODP approach \cite{england99}. The age effect in the first development period is set to zero to make the model identifiable.}
        \label{fig:apc1}
    \end{subfigure}
    \hfill
    \begin{subfigure}[t]{0.49\textwidth}
        \centering\includegraphics[width=\linewidth]{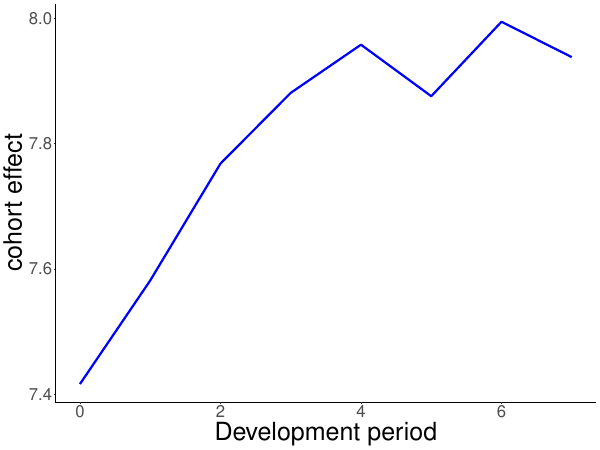}
        \caption{Cohort effect estimated from the GLM approach modeling the claim amount in a cross-classified ODP approach \cite{england99}.}
        \label{fig:apc2}
    \end{subfigure}
\end{figure}

In Figure \ref{fig:paramsestimates} we display the parameters estimates for the three different methods, i.e.,
$\widehat a_j$ for \texttt{clmplus}, $\widehat f_j$ for chain-ladder, and "age-effect" \& "cohort effect" for GLM on claim amounts.
In Figures \ref{fig:clmplus} and \ref{fig:clm} we see, respectively, a decreasing behavior in the age-effect for the \texttt{clmplus} and chain-ladder estimates.
The parameters estimates for the GLM approach on claim amounts are displayed in Figure \ref{fig:apc1} and \ref{fig:apc2}.

\subsubsection{The benefit of adding the cohort and period components}
\label{ss:period}

\begin{figure}[H]
    \centering
        \centering\includegraphics[width=.6\linewidth]{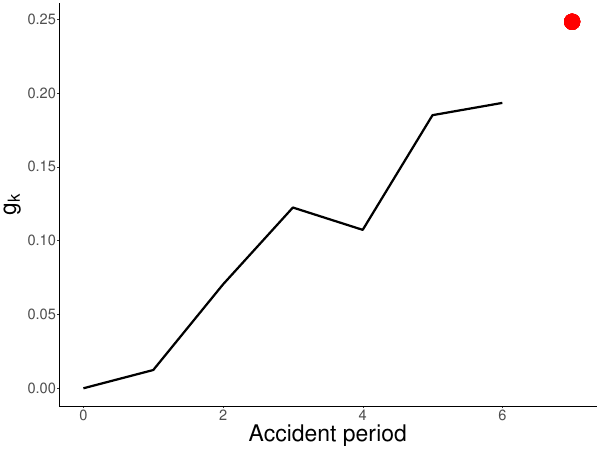}
        \caption{\label{fig:cohortA}Cohort component $g_{k}$ extrapolated with an ARIMA model (1,1,0) with drift, red dot on the \texttt{AutoBI} data.}
\end{figure}

\begin{figure}[H]
    \centering
    \caption{\label{fig:modelsresiduals} Scaled deviance residuals for hazard of the age-model, the age-cohort model, the age-period model and the age-period-cohort model on the \texttt{AutoBI} data.
    }
    \begin{subfigure}[t]{0.49\linewidth}
        \centering\includegraphics[width=\linewidth]{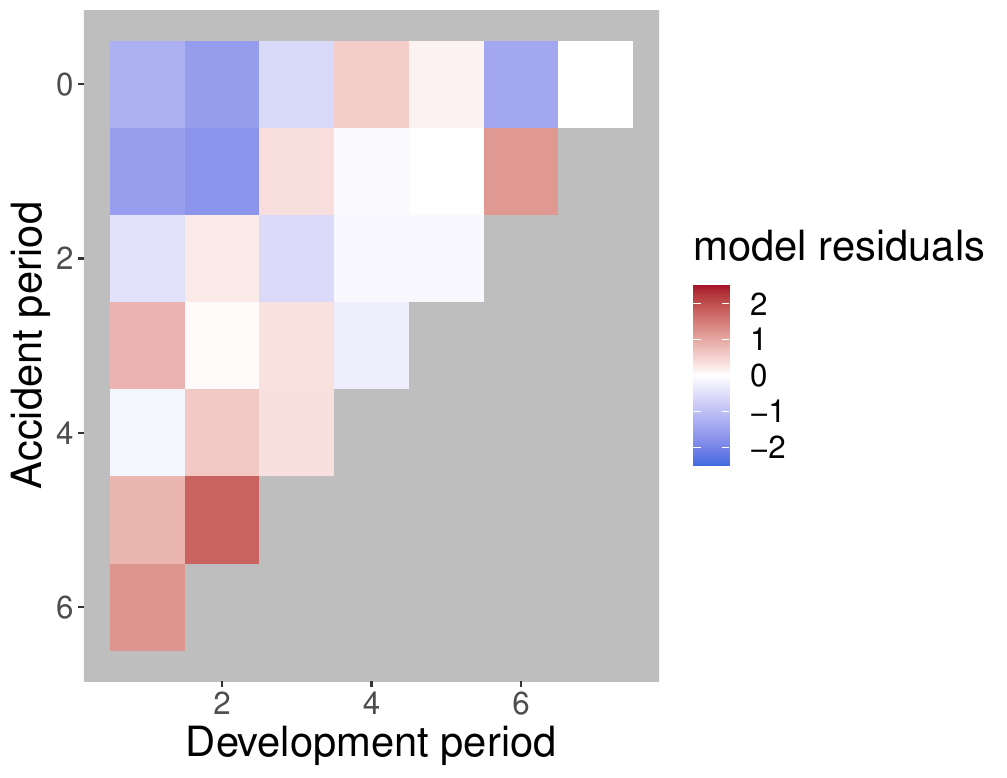}
        \caption{Residuals plot from the age model.}
        \label{fig:resa}
    \end{subfigure}
    \hfill
    \begin{subfigure}[t]{0.49\linewidth}
        \centering\includegraphics[width=\linewidth]{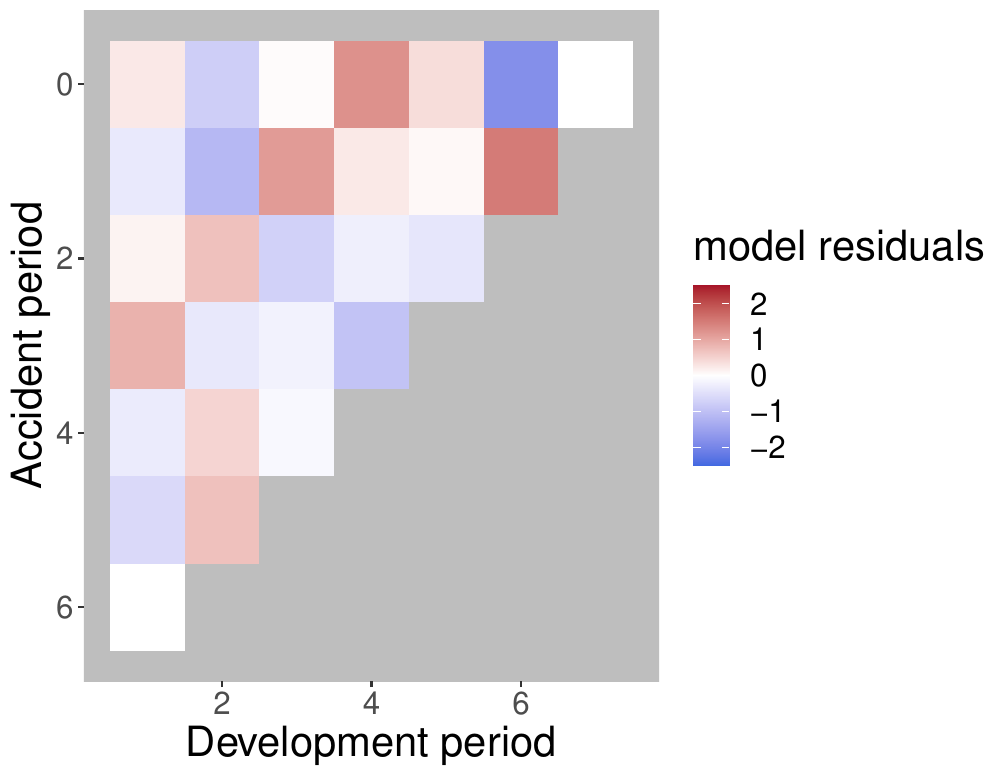}
        \caption{Residuals plot for the age--cohort model.}
        \label{fig:resac}
    \end{subfigure}
    \hfill
    \begin{subfigure}[t]{0.49\linewidth}
        \centering\includegraphics[width=\linewidth]{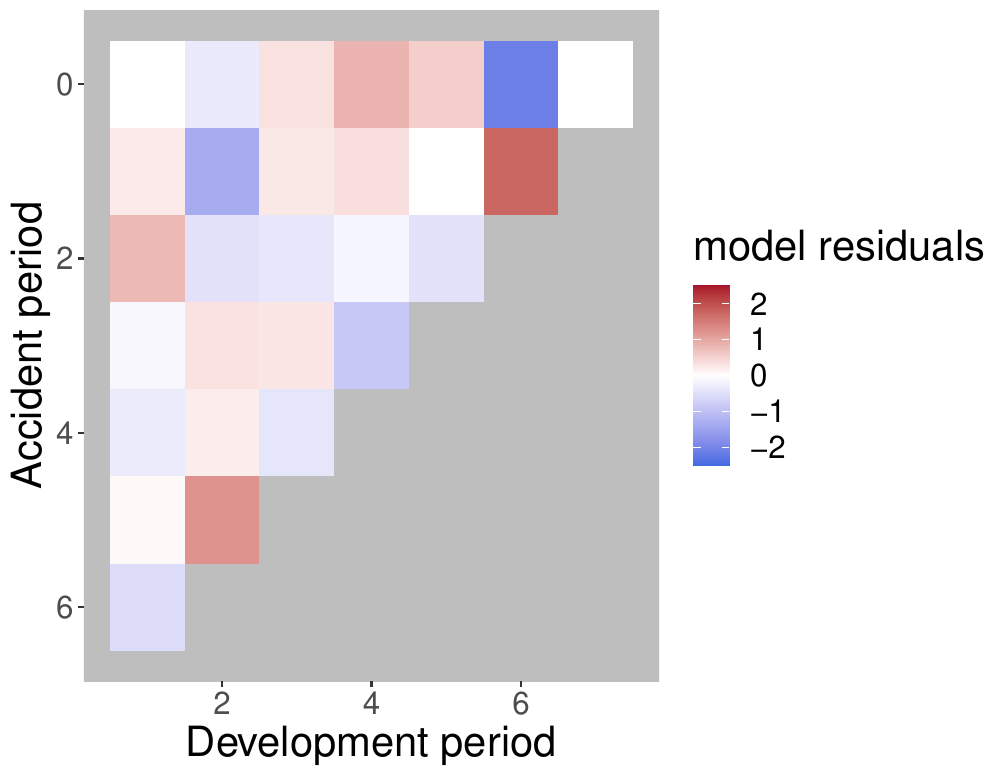}
        \caption{Residuals plot for the age--period--model.}
        \label{fig:resap}
    \end{subfigure}
    \hfill
    \begin{subfigure}[t]{0.49\linewidth}
        \centering\includegraphics[width=\linewidth]{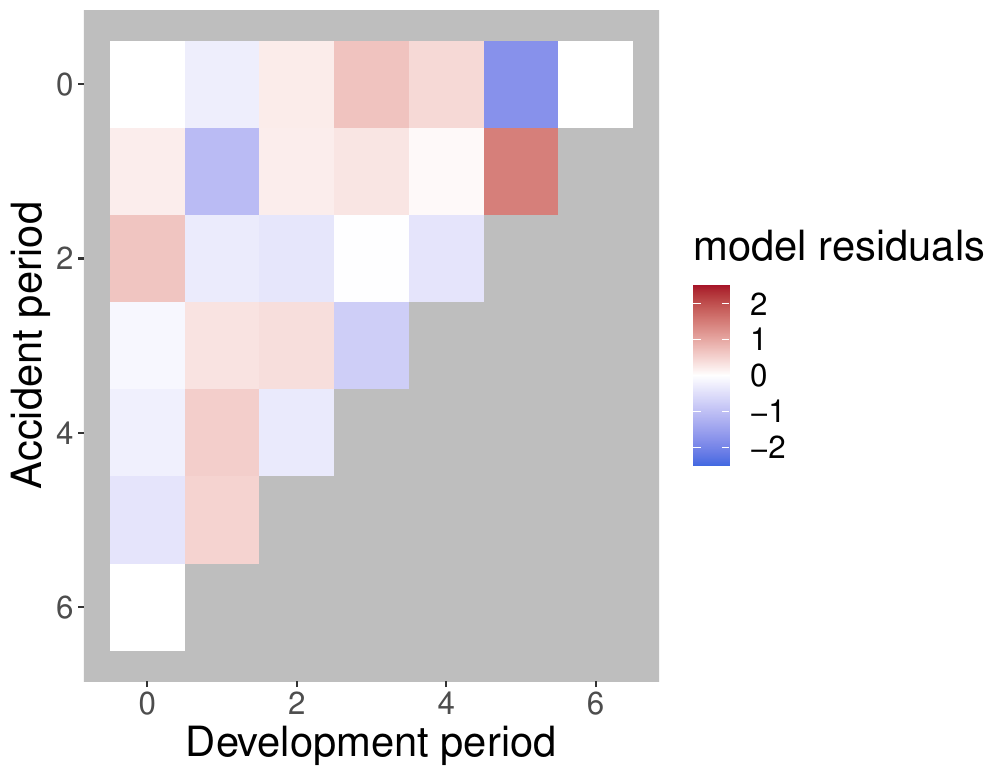}
        \caption{Residuals plot for the age--period--cohort model.}
        \label{fig:resapc}
    \end{subfigure}
    \label{fig:res}
\end{figure}

It is well known and also seen in the previous section that a GLM with claim amount as response and an age component and a cohort component as predictors can replicate the chain-ladder estimates.
We have also seen that we can provide the same results by modeling the claim development via an age component only. This is desirable both from a statistical and a practical perspectives. Indeed, we are able to model less parameters to obtain the chain-ladder reserve and could use the cohort effect as a potential additional improvement to the model fit.
To illustrate this, in Figure \ref{fig:res} we show heat-maps of scaled deviance residuals for the models in Table \ref{tab:presetmodels} fitted to the \texttt{AutoBI} dataset. For the age-only model in Figure \ref{fig:resa}, we identify two residuals clusters  on the heat-map.
Conversely, the estimates obtained from the age-cohort model in Figure \ref{fig:resac} show that with no need of adding a period component, we are able to improve the residuals on the fit.  Remember that one only needs to extrapolate one point if modeling a cohort effect, while one needs to extrapolate $m$ points ahead if modeling a period effect.
The result from the $\widehat g_m$ effect extrapolation using an ARIMA(1,1,0) as described in previous sections is displayed in Figure \ref{fig:cohortA}.

The flexibility of the chain-ladder plus framework also allows to add a period effect to the claim development; for example via an age-period model or an age-period-cohort model. 
For the \texttt{AutoBI} dataset, Figure \ref{fig:resap} suggests that there is no clear improvement on the residuals moving from an age-cohort to the age-period model. By contrast, Figure \ref{fig:resapc} indicates that there is a slight improvement when we add a period component to obtain a full age-period-cohort specification.

However, we observe once again that additional care is required anytime a period model is extrapolated. In particular, one should be careful in choosing the most suitable extrapolation method for the calendar year component $c_{k+j}$: this must be a data driven choice. 

\begin{figure}[H]
    \centering
    \caption{\label{fig:kt} Period component extrapolated for the age-period-cohort hazard model on the \texttt{AutoBI} dataset.}
    \begin{subfigure}[t]{0.45\linewidth}
        \centering\includegraphics[width=\linewidth]{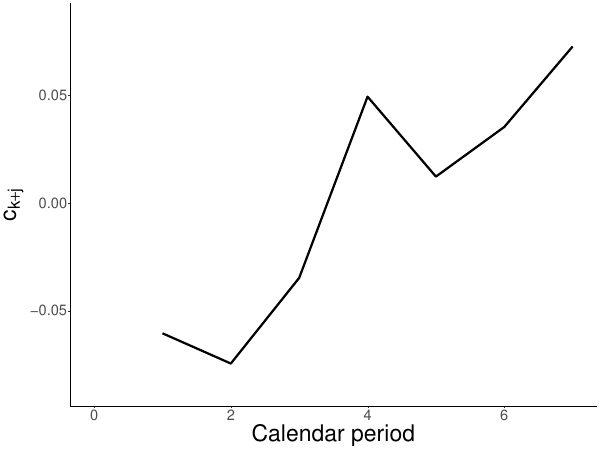}
        \caption{$c_{k+j}$ fitted on the data. Results on the upper triangle.}
        \label{fig:kta}
    \end{subfigure}
    \hfill
    \begin{subfigure}[t]{0.45\linewidth}
        \centering\includegraphics[width=\linewidth]{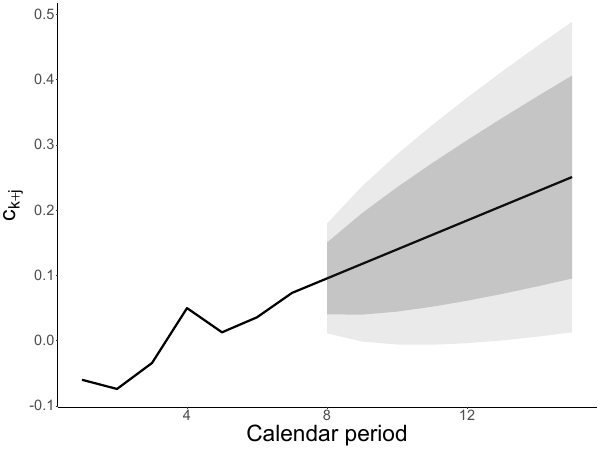}
        \caption{$c_{k+j}$ component extrapolated for the lower triangle. We display the $80\%$ and $95\%$ confidence bounds.}
        \label{fig:extrakt}
    \end{subfigure}
\end{figure}

Figure \ref{fig:kt} shows the $c_{k+j}$ component in an age-period-cohort fit on the upper triangle (Figure \ref{fig:kta}) and the extrapolation in the lower triangle (Figure \ref{fig:extrakt}). For this figure, we fitted the age-period-cohort model 
with the constraints $\sum_k g_{k} =0$ and $ \sum_k k g_{k}=0$.
Furthermore, we assumed here and in the following sections that the period components follow a random walk with drift for forecasting. {\color{black} An actuary might choose a different ARIMA model from the one discussed in Section \ref{ss:effectsextr} based on case-specific data. This may involve implementing one of the model selection criteria for time series analysis, see for example \citeA[ch.~2]{shumway00}.}

Lastly, in Table \ref{tab:hmreserves} we show the different reserves estimates based on \texttt{clmplus} for different cohorts according to the different models we considered in this section. In this example, 
\texttt{ac}, \texttt{ap} and \texttt{apc} produce more similar results compared to the \texttt{a} model. In general, we can see that differences
are more pronounced in the later accident years.

\begin{table}[ht]
\caption{\label{tab:hmreserves} Reserves computed according to different claim development models on the \texttt{AutoBI} data.}
\centering
\begin{tabular}{rrrrr}
  \hline
accident year & a & ac & ap & apc \\ 
  \hline
1 & 0.00 & 0.00 & 0.00 & 0.00 \\ 
  2 & 67.24 & 68.20 & 68.72 & 68.54 \\ 
  3 & 345.19 & 361.77 & 358.22 & 359.35 \\ 
  4 & 940.69 & 1009.65 & 992.50 & 996.34 \\ 
  5 & 2350.86 & 2476.54 & 2503.56 & 2505.20 \\ 
  6 & 4466.77 & 4968.70 & 4845.14 & 5006.93 \\ 
  7 & 9103.24 & 10052.81 & 10229.09 & 10029.15 \\ 
  8 & 14480.44 & 19188.40 & 18377.78 & 19533.02 \\  
  \hline
  Total & 31754.43 & 38126.05 & 37375.01 & 38498.54\\ 
   \hline
\end{tabular}
\end{table}

\subsection{Simulation study mimicing environmental changes}
\label{ss:environment2}

\begin{figure}[H]
        \centering
        \includegraphics[width=10cm]{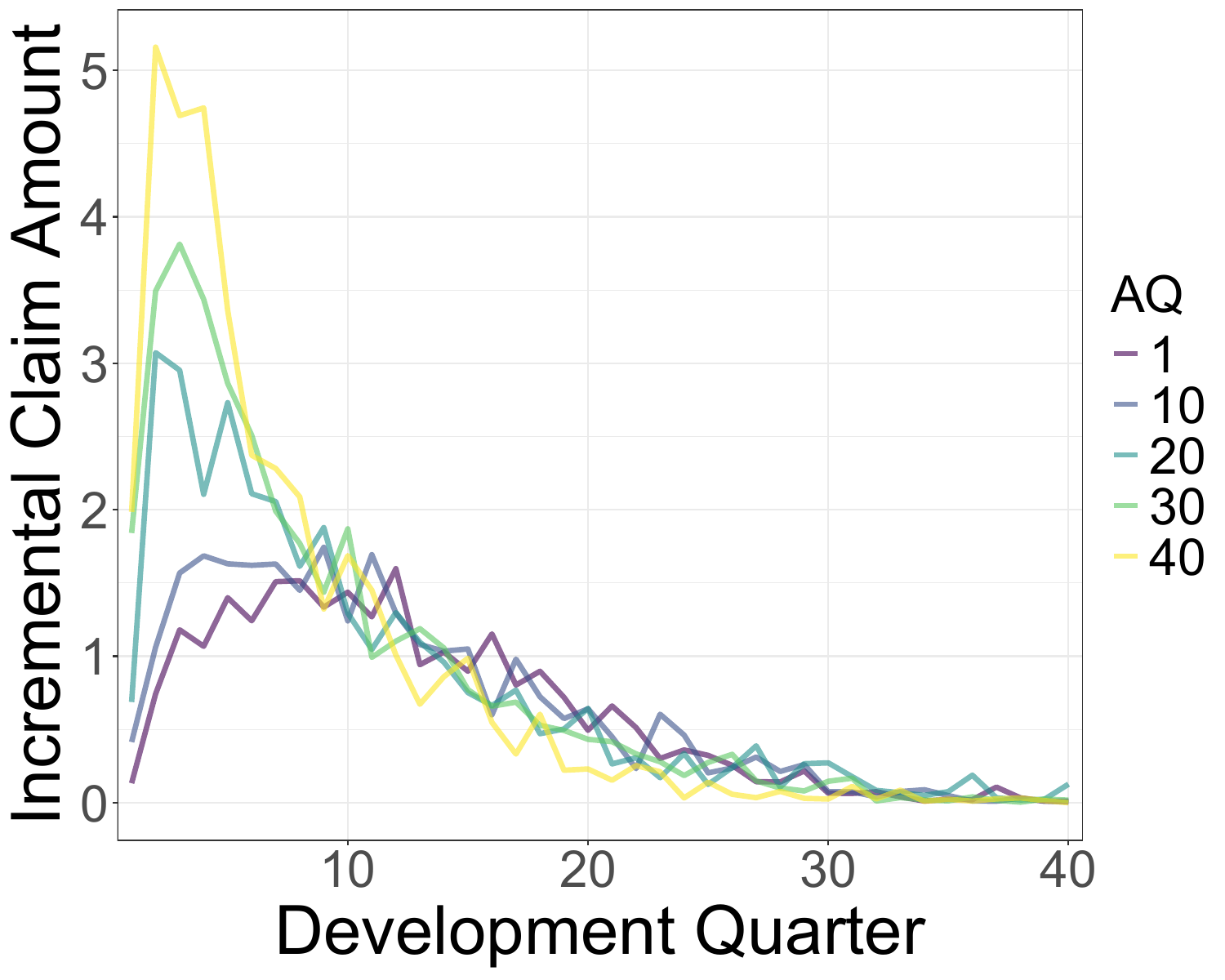}
        \caption{\label{fig:standalone_fig2} We select Accident Quarters $1, 10, 20, 30$ and $40$ for one data set simulated from Environment $2$. We show the incremental claim amount (y-axis) in the different development quarters (x-axis). }
\end{figure}

{\color{black} 

To enhance the strength of our conclusions about the applicability of our models compared to the chain-ladder literature benchmark, we performed a simulation study on data affected by environmental changes. The simulation study is based on  the environments introduced in \citeA{almudafer21}. In this section we consider their Environment $2$, further environments are considered in Appendix \ref{appendix:simulationstudy}. Environment 2 assumes that the development of claims changes from predominantly long tail to predominantly short tail along the accident dimension. \autoref{fig:standalone_fig2}  displays the claim development pattern for different accident quarters from one simulation.
The idea is to mimic an increase in claims processing speed}.

This type of environmental change is also discussed in \citeA{okine22}.
The accuracy of our forecasts across different models is compared using the incidence of the absolute prediction error on the true reserve ($\text{EI}_\text{R}$),

\[
\text{EI}_\text{R} = \left|\frac{\sum_{k+j>m}\widehat X_{kj}}{\sum_{k+j> m} X_{kj}}-1\right|.
\]

The realized future payments $X_{kj}$ with $k=0,\ldots, m$ and $j=1,\ldots, m$ for $k+j>m$ are available from the simulation.  
We run 50 simulations and  we compare the chain ladder model performance to the following three sets of potential models:
\begin{enumerate}[label=(\alph*)]
    \item the \texttt{apc} package family that models the claim amount,
    \item the  \texttt{clmplus} package family that models the claim development, and 
    \item the union of (a) and (b).
\end{enumerate}

For validation and testing, we adopt the approach illustrated in Figure \ref{fig:naicbo}. In particular, on each of the $50$ runs and for each of the three model families (sets (a), (b) and (c)), we first choose the model that minimizes the $\text{EI}_\text{R}$ on the validation data. Secondly, we measure the performance of the selected model on the test data. 
\newline

\begin{figure}
\centering
\includegraphics[width=\linewidth]{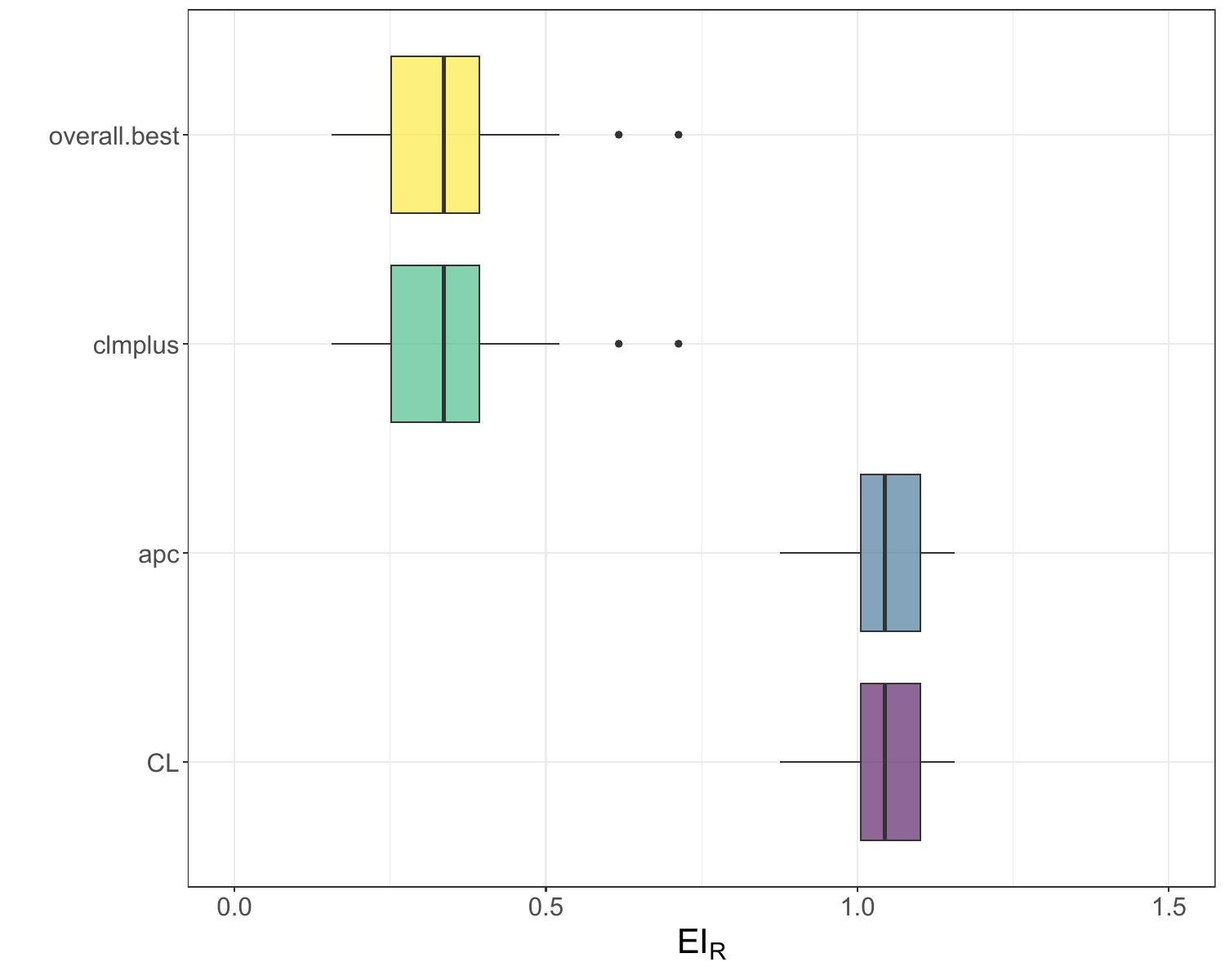}
        \caption{\label{fig:environment2} Models $\text{EI}_\text{R}$ on the test set across the $50$ simulated datasets simulated from Environment $2$. The different model sets include the chain-ladder(violet), \texttt{apc} (blue), \texttt{clmplus} (green) and \texttt{overall.best} (yellow).}
\end{figure}

The results in terms of $\text{EI}_\text{R}$ on the test set are shown in Figure \ref{fig:environment2}. The \texttt{apc} package models (a) seem to perform similarly to the literature benchmark, the chain-ladder (CL in the Figure). Conversely, we find that the \texttt{clmplus} models (b) 
delivers a great improvement in this scenario
having a better $\text{EI}_\text{R}$ performance than (a). 
Comparable performance of set (c) (the union of \texttt{apc} and \texttt{clmplus}) and set (b) (\texttt{clmplus}) indicates that set (a)
(\texttt{apc}) never led to any improvement in this scenario. This is also confirmed  in Table \ref{tab:setcchoices}. 
Here we find  that in the 50 runs the following different models have been chosen in (c):
$1\times$ \texttt{ac} (\texttt{clmplus}), $19\times$ \texttt{apc} (\texttt{clmplus}), $30\times$ \texttt{ap} (\texttt{clmplus}). 

\subsection{Benchmarking on multiple data sets}\label{sec:ModelComparison}

While in Section \ref{ss:autobisuper} we showed a sensible strategy for model selection on a single run-off triangle, within this section we want to assess our model performance on several real datasets.
We gathered $30$ real and  publicly available  run-off triangles from the R packages \texttt{ChainLadder}, \texttt{apc}, \texttt{clmplus} and \texttt{CASdatasets} \cite{casdatasets}. The complete list of the triangles we used is provided in \autoref{tab:sectionfive}, in Appendix \ref{appendix:triangleslist}.
Our objective is to show that practitioners can really benefit from the chain-ladder plus model framework as an additional more versatile tool-box to the already rich literature on claims reserving.
After a short discussion on model selection, we compare the performances of chain-ladder plus models in Table \ref{tab:presetmodels} using the \texttt{clmplus} R-package to the performances of the age-cohort and age-period-cohort GLM based on claim amounts using the \texttt{apc} R-package, as outlined in \citeA{kuang11} and implemented in \citeA{package:apc}.

For model comparison, it is not straightforward to select a good measure of fit.
One issue is that the mean squared error or the mean absolute error would not be comparable across different triangles that exhibit different payments size.
Another issue is related to the time series structure of run-off triangles. 
It is worth noticing again that anytime the claims reserve is computed, the results on the lower triangle are extrapolated. In order to assess the models capability to extrapolate, a reasonable approach is to use the most recent diagonals as test set.
Thus, we will evaluate the performance of different models across diagonals in terms of absolute errors incidence ($\text{EI}_l$) on a selected diagonal $l$,

\[
\text{EI}_l=\left|\frac{\sum_{k+j=l}\widehat X_{kj}}{\sum_{k+j= l} X_{kj}}-1\right|.
\]

{\color{black} To facilitate clear interpretation of the results, we will not discuss the possible performance metrics further. Other metrics than $\text{EI}_l$ are certainly possible and could be defined using expert judgement.}


\subsubsection{Models ranking}
\label{ssec:modelsranking}

In order to rank the models performances, we evaluated on the calendar year $m$ the best performing model in terms of error incidence over the cumulative payments diagonal. An example for a $12\text{x}12$ run-off triangle is provided in Figure \ref{fig:modelsvalmr} in Appendix \ref{appendix:modelselectionsplits}.

\begin{figure}[H]
    \centering\includegraphics[width=18cm]{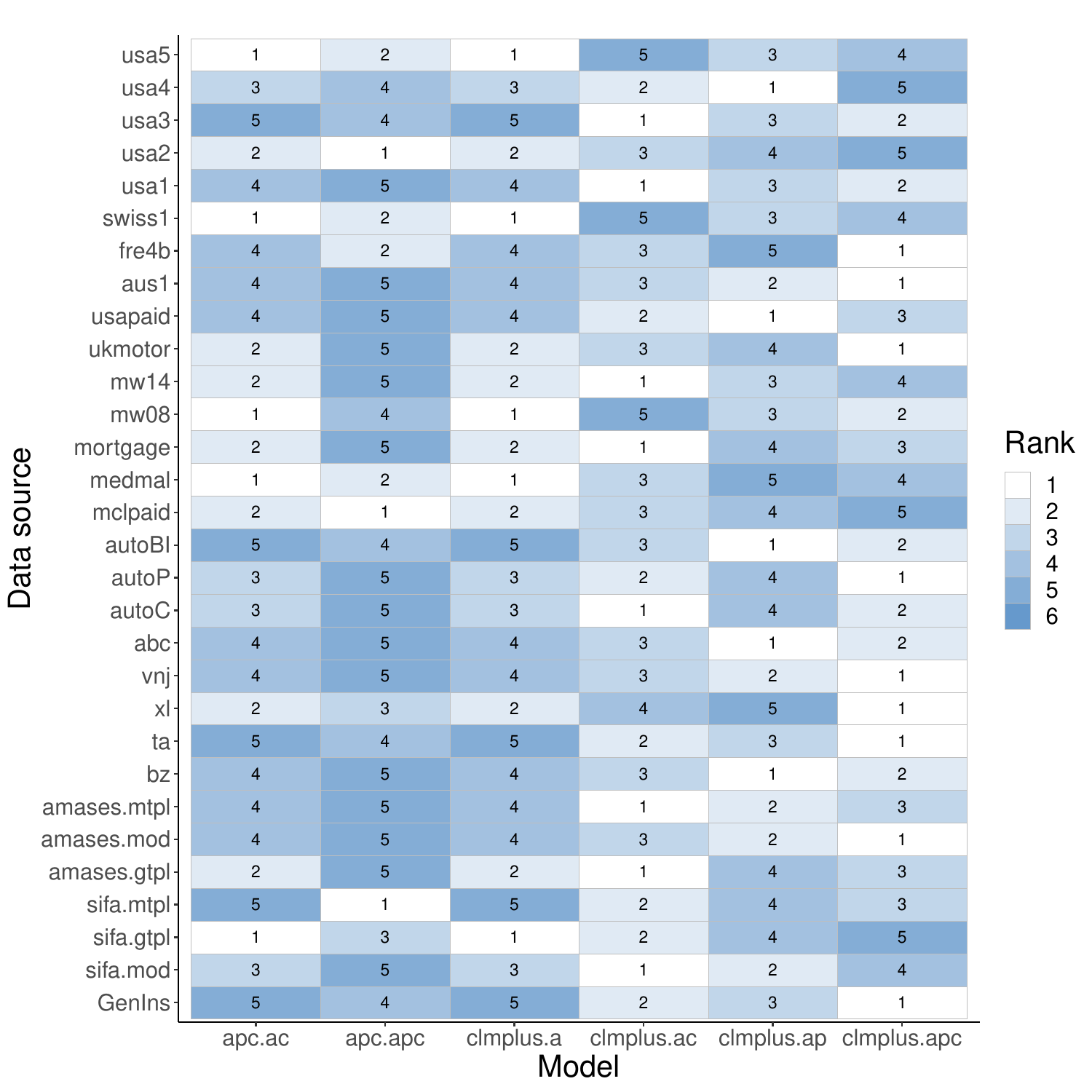}
    \caption{\label{fig:ranks} Models rankings based on the $\text{EI}_l$ performance for the $30$ different real dataset listed in \autoref{appendix:triangleslist}.}
\end{figure}

On each dataset, the models were first trained on the training set, see the blue area in Figure \ref{sfig:valimr}. 
The error incidence was then recorded for each model on the last diagonal, see the red area in Figure \ref{sfig:valimr}. 
The models where then ordered in ascending order: the model with the lowest absolute error incidence was selected to be the best performing. Figure \ref{fig:ranks} shows ranking of models for each of the 30 datasets.




To facilitate comparison, Table \ref{tab:meanranks} shows the mean rank of each model across the 30 different datasets. In Figure \ref{fig:ranks} and Table \ref{tab:meanranks}, it is worth noticing once again that the age-cohort model in \citeA{england99} and the chain-ladder plus age-model yield the same results on every dataset. In addition, the chain-ladder plus models tend to show the best performance, with particular reference to the age-period and age-period-cohort structure. 

\begin{table}[H]
\centering
    \begin{tabular}{rllllll}
  \hline
Model &apc.ac & apc.apc & clmplus.a & clmplus.ac & clmplus.ap & clmplus.apc \\ 
  \hline
 Mean rank & 3.07 & 3.87 & 3.07 & 2.47 & 3.00 & 2.60 \\ 
   \hline
\end{tabular}
\caption{\label{tab:meanranks} Average model rankings based on the $\text{EI}_l$ performance for the different models across all the $30$ data sets in \autoref{appendix:triangleslist}.} 

\end{table}

\subsubsection{Model families comparison}
\label{ssec:bakeoff}

In this section we evaluate whether one would benefit from having at disposal a larger set of models to choose from when carrying out a reserving exercise. To do so, we compare the performance of three sets of potential models on each of the 30 datasets used in the previous section. These sets of potential models are {\color{black} described in Section \ref{ss:environment2}.}

To evaluate the performance of each set we start by splitting the data into training, validation and testing as illustrated in Figure \ref{fig:modelsvalbo} in Appendix \ref{appendix:modelselectionsplits}. Then, for each dataset the best model within the three different model sets is selected based on the validation set, and, finally, the error incidence ($\text{EI}_m$) of this best model is then calculated on the test set.  

\begin{figure}[H]
        \centering\includegraphics[width=.9\linewidth]{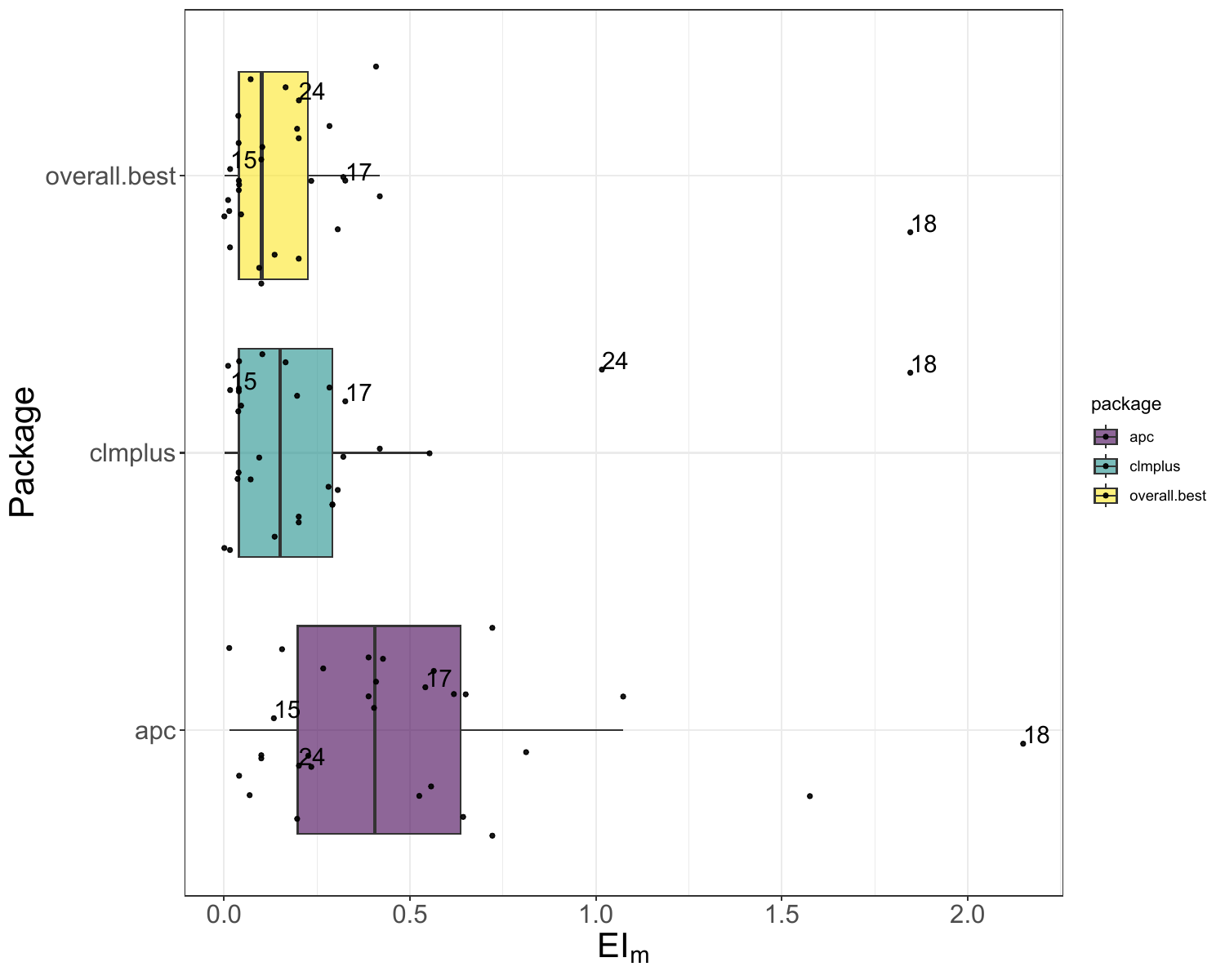}
        \caption{\label{fig:eibakeoff} 
   Box-plot of the models $\text{EI}_m$ on the test set across the $30$ datasets in \autoref{appendix:triangleslist}. On each dataset we selected the best performing model via a validation set from the three families 
\texttt{overall.best}, \texttt{clmplus} and \texttt{apc}. }
\end{figure}

Figure \ref{fig:eibakeoff} shows box plots of the $\text{EI}_m$ for the 30 datasets and each of the three sets of models. 
We find that the \texttt{clmplus} package family (b) has on average a better performance than the \texttt{apc} package family (a). However, more importantly, having all models at disposal (set (c)) allows to further improve the forecasting accuracy,  demonstrating that it is useful to expand the set of models available in the reserving practice toolkit. It is especially reassuring that picking a model via a validation set seems to generally lead to improved one-year-ahead predictions.

For illustrative purposes, we marked data sets $15,17,18, 24$ in Figure \ref{fig:eibakeoff}. In data set $24$, using the best model from the \texttt{apc} family, considerably reduced the $\text{EI}_m$ compared to picking the best model from the \texttt{clmplus} family. The opposite occurs for data set $15$.

\subsection{Case study on the National Association of Insurance Commissioners (NAIC) data}
\label{sec:naic}

In this Section, we consider a case study using historical fully developed market data. The data is publicly available, and it was pulled from \textit{Schedule P - Analysis of Losses and Loss Expenses} in the National Association of Insurance Commissioners (NAIC) database \cite{naic}. The data aggregates run-off triangles of six lines of business for all U.S. property casualty insurers. The triangle data contain the claims of accident years $1988$ to $1997$ with $10$ years development years. In our application, we will consider the following lines of business: Commercial Auto, 
Medical Malpractice, Other Liability, Private Passenger Auto, Workers Compensation.

We split each data set into training validation and testing as illustrated in Figure \ref{fig:naicbo} in Appendix \ref{appendix:modelselectionsplits} and repeat, on the NAIC data, the comparison of the three model families illustrated {\color{black}in Section \ref{ss:environment2}}. In particular, we consider the data from the calendar period $k+j=m$ as the validation set and the data from the subsequent periods $k+j>m$ as the test set (lower triangle). 
Having now the lower triangle available, allows us to use the future losses to measure the accuracy of our forecasts using the incidence of the absolute prediction error on the true reserve ($\text{EI}_\text{R}$).

\begin{figure}[ht]
\centering\includegraphics[width=.9\linewidth]{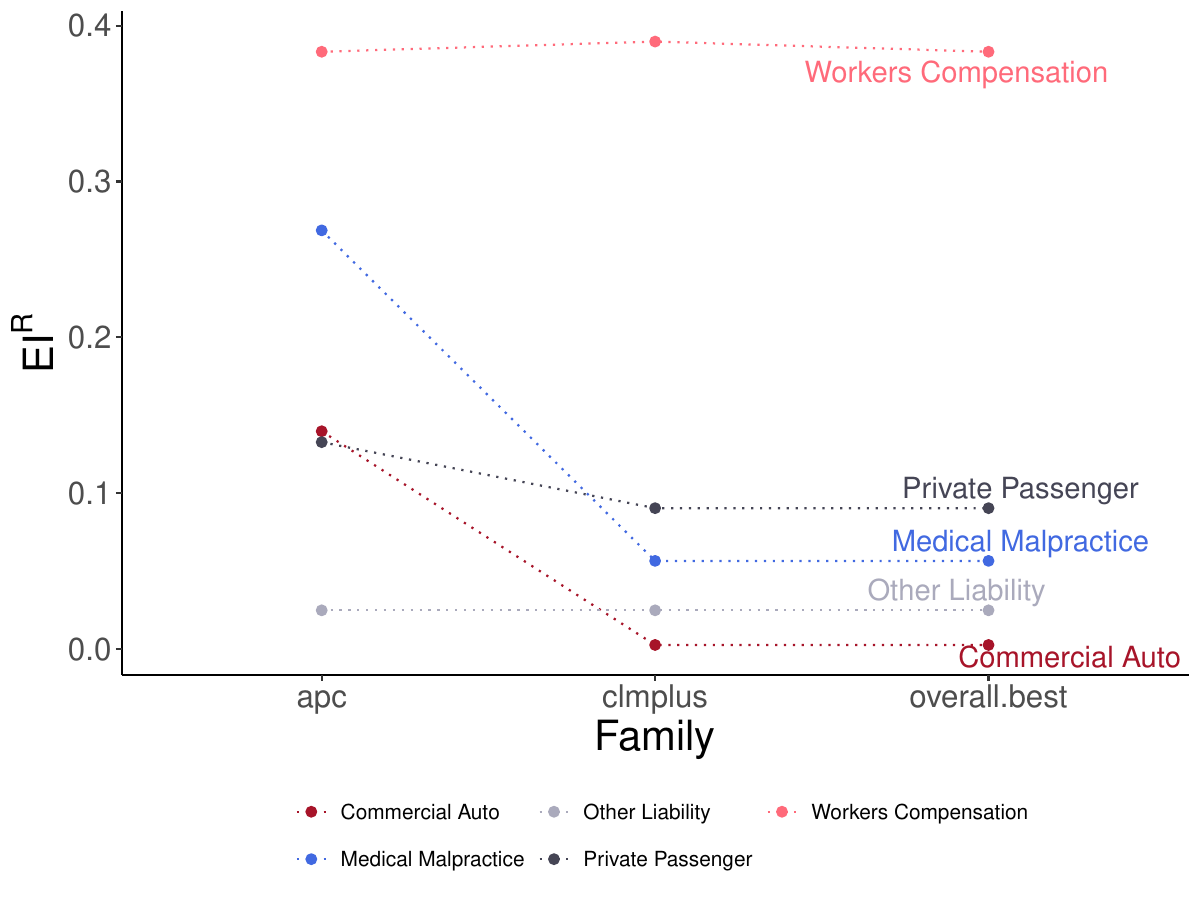}
        \caption{\label{fig:bakeoffreserve} 
   Models $\text{EI}_{\text{R}}$ on the test set across the NAIC datasets. On each dataset we selected the best performing model via a validation set from the three families 
\texttt{overall.best}, \texttt{clmplus} and \texttt{apc}. The dotted line connects error measurements on the same data set. }
\end{figure}

We displayed the $\text{EI}_\text{R}$ for the different lines of business in Figure \ref{fig:bakeoffreserve}. The plot shows, for each family of model the $\text{EI}_\text{R}$ on the test of the best performing model among each family for each data set. Interestingly, the union of the \texttt{apc} and \texttt{clmplus} families of models always obtains a lower prediction error on the reserve compared to using only one of the two families. This indicates that the validation procedure of picking the best model by evaluating the performance on the last diagonal works well in the cases considered.

The results also show that, in the inspected data sets, the models from the \texttt{clmplus} family generally perform better than the models from the \texttt{apc} family in terms of $\text{EI}_\text{R}$. The only exception is the Line Workers Compensation where we select the apc model from the \texttt{apc} family. In the Line Other Liability, we select the Age Model (Chain Ladder). The performance measurements of the chosen models on the test set are shown, for each Line of Business and each family in Table \ref{tab:naic}.

\begin{table}[ht]
\centering
\begin{tabular}{l|rr|r}
  \hline
Line of Business & \texttt{apc} & \texttt{clmplus} & \texttt{overall.best} \\ 
  \hline
Commercial Auto & 0.140 (ac) & 0.003 (ac) & 0.003 (ac, \texttt{clmplus})\\ 
Medical Malpractice & 0.269 (ac) & 0.057 (ap) & 0.057 (ap, \texttt{clmplus})\\ 
Other Liability & 0.025 (ac) & 0.025 (a) & 0.025 (a, \texttt{clmplus})\\ 
Private Passenger & 0.133 (ac)& 0.090 (ap)& 0.090 (ap, \texttt{clmplus})\\ 
Workers Compensation & 0.383 (apc) & 0.390 (ac)& 0.383 (apc, \texttt{apc})\\ 
   \hline
\end{tabular}
\caption{\label{tab:naic} For each Line of Business of the NAIC data (column one) and each models set (columns two to four), we show the $\text{EI}^\text{R}$ on test set of the model that we pick on the validation sets. }
\end{table}

\section{Conclusions}

In this paper, we introduced a framework to model the claims development via a GLM.
This is in contrast to existing GLM based reserving literature which models the claim amount.
We showed that modeling the claim development via an age model replicates chain-ladder's predictions.
The simplicity of the model furthermore invites very naturally to consider model extensions beyond the simple age model.  In the data sets considered,  including additional effects such as calendar effects and cohort effects often improved the fit. Especially adding a cohort effect seemed to often be beneficial.
Interestingly, adding a period effect when modeling the claim amount did often not lead to any improvement. This is possibly due to the fact that inflation is often adjusted for before being aggregated  into run-off triangles.
But there are exceptions and our data study demonstrated that the best results are obtained by having both 
claim amount models and claim development models as well as various structures at one's disposal.
\newline
An interesting extension of the present work would be to model the development of the claim frequency based on individual data and within a recent granular reserving frameworks, see e.g. \citeA{pigeon13}, \citeA{antonio14}, \citeA{baudry19}, \citeA{lopez19}, \citeA{lopez21}, \citeA{okine22}, \citeA{delong22}, \citeA{crevecoeur22}. 
The modeling of the claim development could be done with recent machine learning methods such as tree based models and neural networks which have shown to handle the inclusion of a number of covariates well.

{\color{black} Lastly, our model is not capable of estimating the distribution or standard error of the reserve but only provides point estimates.
In the case of payment triangles, one would for example need additional assumptions on the distribution of individual claim sizes.
While this is certainly possible, the estimated standard errors would heavily depend on the specific assumptions made. 
It would be interesting to investigate in the future whether simple summary statistics from individual claims data could help with this task leading to more trustworthy uncertainty quantification of the reserve than current techniques based on purely aggregated data.
}

    



\bibliography{main.bib}


\appendix

\section{Granular model formulation}\label{appendix:le}

\subsection{The model}
Given a cut-off-date, we have observed $n$ claim payments.
For every payment,  we are given the time delay from accident until payment, $T_i$, and payment size $Z_i$, $i=1,\dots,n$.
We make the following assumption. 
\begin{enumerate}
    \item[{[M1]}] All payments are independent.
\end{enumerate}
Assumptions [M1] is rather strong but is made to simplify the mathematical derivations and we conjecture that it can be relaxed by considering weak dependency between payments.  
As pointed out in \citeA{hiabu17} and \citeA{bischofberger2020continuous}, 
statistical inference on  $(T_i, Z_i)$ is not directly feasible.
We only observe  $(T_i, Z_i)$ when the payment date is before the cut-off-date. Therefore, by design it holds
\[
 T_{i}\leq \textit{cut-off-date} -U_i,
\]
where $U_i$ is the accident date of claim $i$. 
Hence, by not following every policy, we are exposed to a
right-truncation problem instead of a right-censoring problem.
In the sequel, for notational convenience, we parameterize the dates such that $ \textit{cut-off-date} = \mathcal T$ which yields $T_{i}\leq \mathcal T -U_i$.
A solution to the right-truncation problem is to reverse the time of the counting process leading to a tractable left-truncation problem \cite{Ware:DeMets:76}.
To this end we consider the development-time reversed counting processes
\[
N_i(t)= I(t\geq \mathcal T- T_{i}),
\]
each with respect to the filtration
\begin{equation*}
\mathcal F_{it}=\sigma \left( \bigg\{ T_i -t \leq s:\ s\leq t\bigg\} \cup \bigg\{    U_i \bigg\} \cup \mathcal N\right),
\end{equation*}
satisfying the \textit{usual conditions} \cite[p.~60]{andersen93}, and where $\mathcal N$ is the set of all zero probability events. 
Notice that the right truncation problem has transformed into a left truncation problem:
\[
T_{i}^R\geq  U_i, \quad T_{i}^R= \mathcal T- T_{i}.
\]
Additionally and in contrast to  \citeA{hiabu17} and \citeA{bischofberger2020continuous}, we allow $T_i$ to depend on $Z_i$.
Then, under mild regularity conditions, the intensity process of $N_i^R$ is
\[
\lambda_i(\mathcal T-t|u)=\lim_{h \downarrow 0} h^{-1}E\left[  N_i\left\{(\mathcal T-t+h)-\right\}- N_i(\mathcal T-t-)|\ \mathcal F_{i,(\mathcal T-t)-}\right]
=\alpha^R(t|U_i)Y_i^R(t),
\]
where 
\begin{align*}
\alpha^R(t|u)&=\lim_{h \downarrow 0} h^{-1}{P}\left(T_i\in(t-h,t]|\ Y_i^R(t)=1, U_i=u\right), \\
Y_i^R(t)&=I(T_i \leq t < \mathcal T-U_i),
\end{align*}
This structure is called Aalen's multiplicative intensity model 
\cite{aalen78}, and enables nonparametric estimation and inference 
on the deterministic factor $\alpha^R(t|u)$.

Let $Z_i$ denote the payment size of claim $i$
and consider the process $N^\ast_i(t)=Z_i N_i(t)$. 
Under mild regularity conditions, 
it is straightforward to see that 
\begin{align*}
 \lambda^\ast_i(\mathcal T-t)&=\lim_{h \downarrow 0} h^{-1}E\left[   N^\ast_i\left\{(\mathcal T-t+h)-\right\}- N^\ast_i(\mathcal T-t-)|\ \mathcal F_{i,(\mathcal T-t)-}\right]=\mathcal R_n(t)  \alpha^{\ast,R}(t|U_i) Y_i^{\ast,R}(t),\\
 \alpha^{\ast,R}(t|u)&=
\frac{  E[Z_1 | T_1=t, U_1=u]}{ E\left[Z_1 | \ T_1 \leq t, U_1=u\right]} \alpha^R(t|u),\\
Y_i^{\ast,R}(t)&=Z_i Y^R_i(t),\\
\mathcal R_i(t,u)&= \frac{E\left[Z_1 | \ T_1 \leq t, U_1=u\right]} {Z_i}. \\
\end{align*}

\subsection{Introducing a grid system} \label{grid_system}
We assume that observations are aggregated in periods of size $\delta$. For notations convenience, we assume that the same aggregation level $\delta$ is chosen for both development  direction, $T_i$ and accident direction, $U_i$ and furthermore that $m=\mathcal T/\delta$ is an integer.
We then work on a equi-distant grid $t_0=0,\dots, t_{m+1}=\mathcal T$ and $u_0=0,\dots, u_{m+1}=\mathcal T$, with $t_j-t_{j-1}=u_k-u_{k-1}=\delta, j,k=0,\dots,m$.
In classical run-off triangles we only have information on the sum of payments falling in a parallelogram
\[
\mathcal{P}_{kj}= \{(t,u): t_j+u_k-u\leq t \leq t_{j+1}+u_k-u; u\in [u_k,u_{k+1} ), t\geq 0\},
\]
i.e., the individual observations $(Z_i,T_i,U_i), i=1,\dots n$ are compressed into observations
$X_{kj};\ j,k=0,\dots,m$ with
\[
X_{kj}= \sum_i Z_i \int
I\left((s,U_i)\in \mathcal{P}_{kj}\right) dN_i(s).
\]

To estimate the development in the run-off triangle, we are interested in the exposure-weighted average hazard on the parallelogram:
\[
\mu_{kj}=\frac{\delta \int_{\mathcal{P}_{kj}} \alpha^{\ast,R}(s|u)p_U(u)\gamma(s,u) ds du} {\int_{\mathcal{P}_{kj}} p_U(u)\gamma(s,u) ds},
\]
where $\gamma(s,u)=E[Y_i^{\ast,R}(s) | U_i=u]$ and $p_U$ is the marginal density of $U_i$.
Under regularity conditions, $\mu_{kj}$ is the asymptotic limit of

\[
\frac{\delta \sum_i\int I\left((s,U_i)\in \mathcal{P}_{kj}\right) \alpha^{\ast,R}(s|U_i) Y_i^{\ast,R}(s) ds} {\sum_i\int I\left((s,U_i)\in \mathcal{P}_{kj}\right)  Y_i^{\ast,R}(s) ds}.
\]

Hence, if observations on the individual level would be available, then
a natural estimator of
 $\mu_{kj}$ is
 
\begin{align*}
\quad & \frac{\sum_i Z_i\int I\left((s,U_i)\in \mathcal{P}_{kj}\right) dN_i(s)} {\delta^{-1}\sum_i\int I\left((s,U_i)\in \mathcal{P}_{kj}\right)  Y_i^{\ast,R}(s) ds}\\
&=\frac{X_{kj}} {\delta^{-1}\sum_i\int I\left((s,U_i)\in \mathcal{P}_{kj}\right)  Y_i^{\ast,R}(s) ds}\\
&=\frac{X_{kj}} {\sum_{l<j} X_{kl}+ \delta^{-1}\sum_{i: I\left((T_i,U_i)\in \mathcal{P}_{kj}\right)=1} \int I\left((s,U_i)\in \mathcal{P}_{kj}\right)  Y_i^{\ast,R}(s) ds}.
\end{align*}
However, if data is given in the form of a run-off triangle, then the second summand in the denominator is not observed and needs to be approximated.
Note that the expression $\delta^{-1}\sum_{i: I\left((T_i,U_i)\in \mathcal{P}_{kj}\right)=1} \int I\left((s,U_i)\in \mathcal{P}_{kj}\right)  Y_i^{\ast,R}(s) ds$ can be written as 
$\eta X_{kj}$ with $\eta \in [0,1]$.
Assuming that $(T,U)$ is uniformly distributed if $(T,U)$ is conditioned on the parallelogram $\mathcal{P}_{kj}$, an unbiased estimator is given by
$0.5 X_{kj}$. To see this note that
for those payments that occurred in the parallelogram $(j,k)$, i.e., 
$I\left((T_i,U_i)\in \mathcal{P}_{kj}\right)=1$, the  length of the  crossing from entering the parallelogram until payment, i.e., in expected terms is equal to $0.5\delta$:

\begin{align*}
\delta^{-2}\int_0^{\mathcal T} \int_0^{\mathcal T} 
 I\left((s,u)\in \mathcal{P}_{kj}\right) 
 \left(t_{j+1}-t-u+u_k \right)\ ds du = \frac{1}{2} \delta
\end{align*}

Hence, we propose to estimate $\mu_{kj}$ via
{\color{black}
\[
\widetilde \mu_{kj}=\frac{X_{kj}} {\sum_{l<j} X_{kl}+ \frac 1 2 X_{kj}}=\frac{X_{kj}}{E_{kj}}.
\]
}
Note that for $k=0,\dots,m; j=1,\dots m$, chain-ladder's individual development factors  are given by
{\color{black}
\[
\widetilde f_{kj}=\frac{\sum_{l\leq j}X_{kl}} {\sum_{l<j} X_{kl}},
\]

}
such that

{\color{black}
\[
\widetilde f_{kj}= \frac{2+\widetilde \mu_{kj}}{2-\widetilde \mu_{kj}}.
\]
}





\section{The relationship between \texorpdfstring{$\widehat f_{kj}$ and $\widehat \mu_{kj}$}{fkj and mkg}}
\label{appendix:fkjmukj}

In this section, we prove the relationship between individual development factors and the estimator for the average reverse time hazard rate,

{\color{black}
\[
\widetilde f_{kj}= \frac{1+(1-\eta) \widetilde \mu_{kj}(\eta)}{1- \eta \widetilde\mu_{kj}(\eta)},
\]
}
with $k=0, \ldots, m$, $j= 1, \ldots, m$, and $\eta\in [0,1]$. 
\newline 

The chain ladder individual development factors are defined as 

{\color{black}
\begin{align}
\label{eq:fkjproof}
\widetilde f_{k,j} &= \frac{\sum_{l\leq j}X_{k,l}}{\sum_{l<j}X_{k,l}} = 1 + \frac{X_{kj}}{\sum_{l<j}X_{k,l}},
\end{align}
}

and the estimator for the average (reverse time) hazard rate is 

{\color{black}
\begin{equation}
\label{eq:mukjproof}
    \widetilde \mu_{k,j}(\eta) = 1 + \frac{X_{kj}}{\sum_{l<j}X_{k,l}+\eta X_{kj}}.
\end{equation}
}
We can rewrite Equation \eqref{eq:mukjproof} as  

{\color{black}
$$
\sum_{l<j}X_{k,l} = \frac{(1-\eta \widetilde \mu_{k,j}(\eta))X_{kj}}{\widetilde \mu_{k,j}(\eta)}.
$$

}

By plugging in $\sum_{l<j}X_{k,l}$ into \eqref{eq:fkjproof}, we obtain

{\color{black}

$$
\widetilde f_{k,j} = 1 + \frac{\widetilde \mu_{k,j}(\eta)}{1-\eta \widetilde \mu_{k,j}(\eta)}.
$$
}

Then, {\color{black} $\widetilde f_{k,j}$} can then be simplified to 

{\color{black}
$$
\widetilde f_{k,j} =  \frac{1 +(1-\eta)\widetilde \mu_{k,j}(\eta)}{1-\eta \widetilde \mu_{k,j}(\eta)}.
$$
}

\section{The age-model and the chain ladder model}
\label{appendix:agecl} 

In this section we prove that the age-model can replicate the chain ladder development factors for any choice of the parameter $\eta$.  In the age-model, we assume for $k=0,\dots,m$ and $j=1, \ldots, m$, $\mu_{kj}= a_j$. Furthermore, the entries $X_{kj}$ are assumed independent given $E_{kj}$ and  for $k=0,\dots,m; j=1,\dots, m,$ they  follow a Poisson distribution. The log-likelihood is

\begin{align*}
    l\left(a_1,\dots a_m \mid X_{kj}, E_{kj} , j=1,\dots, m; j+k\leq m\right) \propto \sum_{j, k}  X_{kj} \log( a_j E_{kj}) - E_{kj}a_j.
\end{align*}

Under these assumptions, it follows from the first order condition that we obtain the following estimator for the average reverse time hazard rate

\begin{align*}
  \widehat a_j =\frac{\sum_{k} X_{kj}}{\sum_{k} E_{kj}}.
\end{align*}

The individual development factors are then

\begin{align*}
\widehat f_{j}= \frac{1+(1-\eta) \widehat a_{j}}{1- \eta \widehat a_{j}}.
\end{align*}

We drop the index $k$ with $k=0,\ldots,m$ as there is no dependency on the accident period. We can rewrite $\widehat f_j$ as 

\begin{align*}
  \widehat f_j &=\frac{\sum^{m-j}_{k=0} \sum_{l<j} X_{kl}+\eta\sum^{m-j}_{k=0} X_{kj}+ \sum^{m-j}_{k=0} X_{kj}-\eta\sum^{m-j}_{k=0} X_{kj}}{\sum^{m-j}_{k=0} \sum_{l<j} X_{kl}+\eta\sum^{m-j}_{k=0} X_{kj}-\eta\sum^{m-j}_{k=0} X_{kj}}
\end{align*}

which simplifies to the chain ladder development factors 

$$
\widehat f_j =\frac{\sum^{m-j}_{k=0} \sum_{l\leq j} X_{kl}}{\sum^{m-j}_{k=0} \sum_{l<j} X_{kl}}
$$

\section{AutoBI dataset}
\label{appendix:autobi}

\begin{table}[H]
\centering
\caption{Cumulative paid claims of a portfolio of automobile bodily injury liability for an experience period of 1969 to 1976.}
\begin{tabular}{rrrrrrrrr}
  \hline
 & 0 & 1 & 2 & 3 & 4 & 5 & 6 & 7 \\ 
  \hline
0 & 1904.00 & 5398.00 & 7496.00 & 8882.00 & 9712.00 & 10071.00 & 10199.00 & 10256.00 \\ 
  1 & 2235.00 & 6261.00 & 8691.00 & 10443.00 & 11346.00 & 11754.00 & 12031.00 &  \\ 
  2 & 2441.00 & 7348.00 & 10662.00 & 12655.00 & 13748.00 & 14235.00 &  &  \\ 
  3 & 2503.00 & 8173.00 & 11810.00 & 14176.00 & 15383.00 &  &  &  \\ 
  4 & 2838.00 & 8712.00 & 12728.00 & 15278.00 &  &  &  &  \\ 
  5 & 2405.00 & 7858.00 & 11771.00 &  &  &  &  &  \\ 
  6 & 2759.00 & 9182.00 &  &  &  &  &  &  \\ 
  7 & 2801.00 &  &  &  &  &  &  &  \\ 
   \hline
\end{tabular}
\label{tab:autobi}
\end{table}

\section{Model selection splits}
\label{appendix:modelselectionsplits}

The data split into training and validation for the models ranking section is displayed in Figure \ref{fig:modelsvalmr} below.

\begin{figure}[H]
    \centering
    \caption{\label{fig:modelsvalmr} 
   The last diagonal is removed from the triangle and it is used as a validation set to calculate the models rank.}
    \hfill
    \begin{subfigure}[t]{0.49\linewidth}
        \centering\includegraphics[width=1\linewidth]{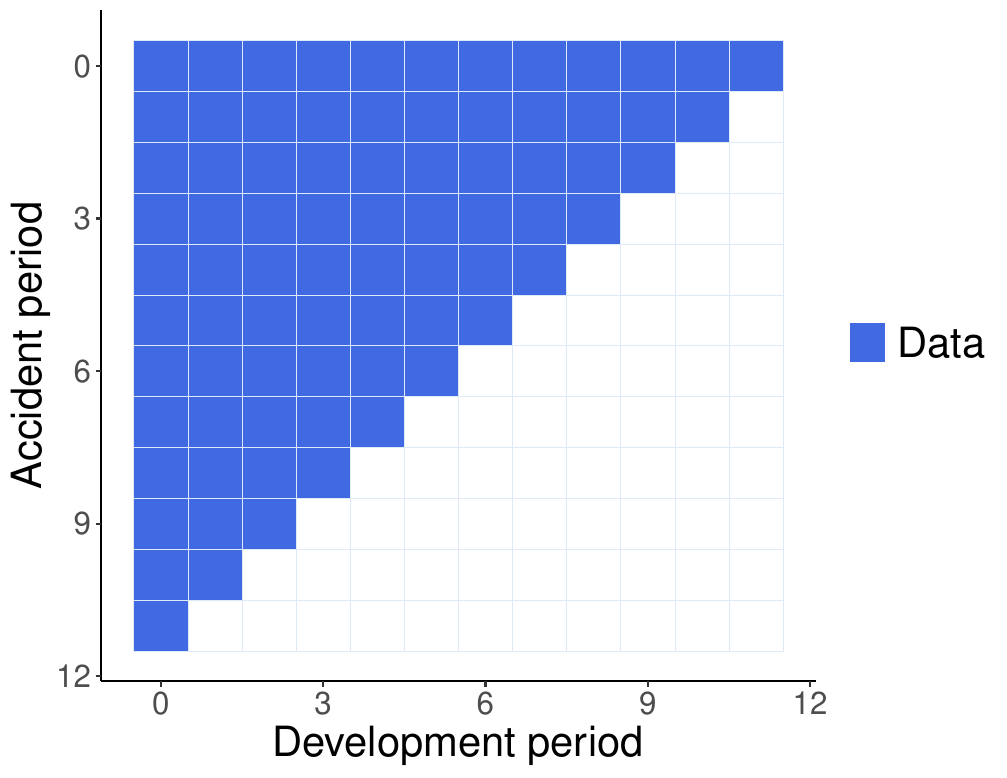}
        \caption{\label{sfig:startingTmr}The starting triangle.}
        
    \end{subfigure}
    \hfill
    \begin{subfigure}[t]{0.49\linewidth}
        \centering\includegraphics[width=\linewidth]{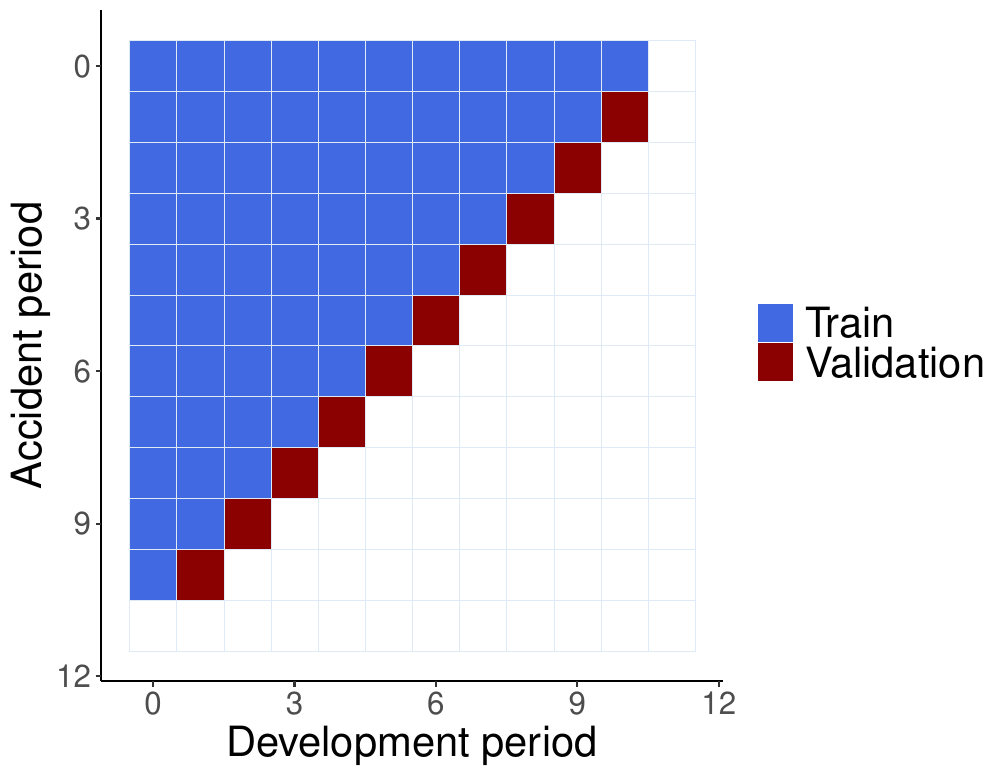}
        \caption{\label{sfig:valimr}The triangle is divided into validation set and training set.}
        
    \end{subfigure}

\end{figure}

\begin{figure}[H]
    \centering
    \caption{\label{fig:modelsvalbo} 
   The last two diagonals are removed from the triangle and they are used as a test set to evaluate the models performance.}
    \hfill
    \begin{subfigure}[t]{0.49\linewidth}
        \centering\includegraphics[width=\linewidth]{figures/full12x12.pdf}
        \caption{The starting triangle.}
        \label{fig:startingTbo}
    \end{subfigure}
    \hfill
    \begin{subfigure}[t]{0.49\linewidth}
        \centering\includegraphics[width=\linewidth]{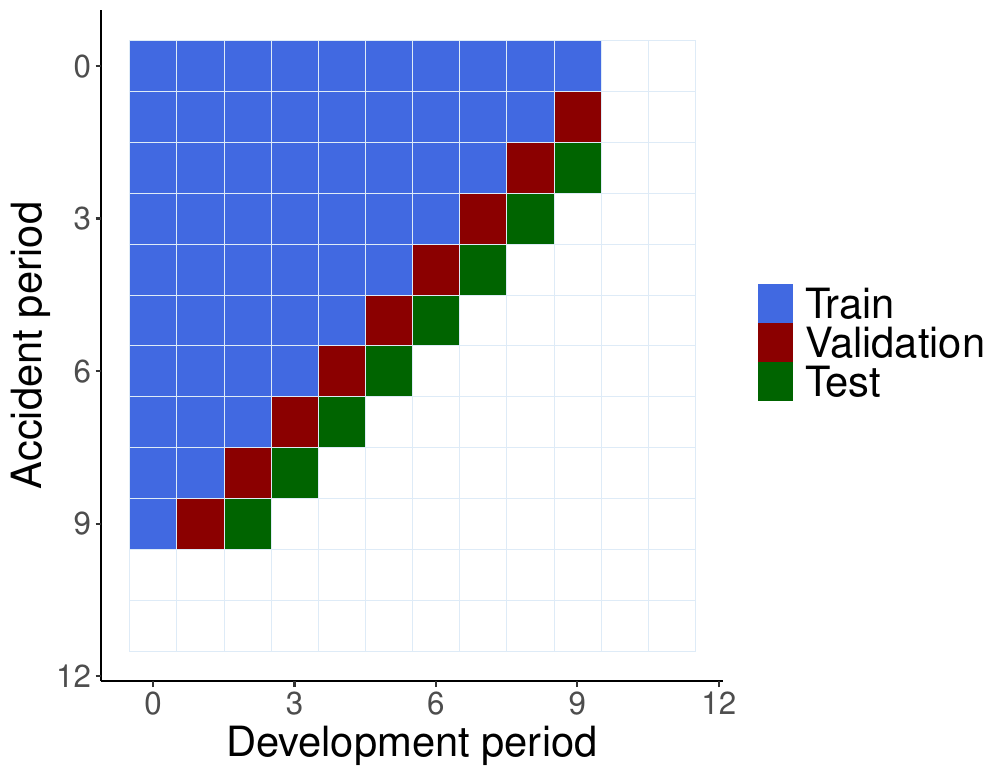}
        \caption{\label{sfig:valibo} The triangle is divided into training set, validation set and test set.}
        
    \end{subfigure}

\end{figure}

Similarly, the data split into training, validation and test used in Section \ref{ssec:bakeoff} is shown in Figure \ref{fig:modelsvalbo}.

\begin{figure}[H]
    \centering
    \caption{\label{fig:naicbo} 
   The last diagonals is removed from the triangle and the lower triangle is used as a test set to evaluate the models performance.}
    \hfill
    \begin{subfigure}[t]{0.49\linewidth}
        \centering\includegraphics[width=\linewidth]{figures/full12x12.pdf}
        \caption{The starting triangle.}
        \label{fig:startingTbonaic}
    \end{subfigure}
    \hfill
    \begin{subfigure}[t]{0.49\linewidth}
        \centering\includegraphics[width=\linewidth]{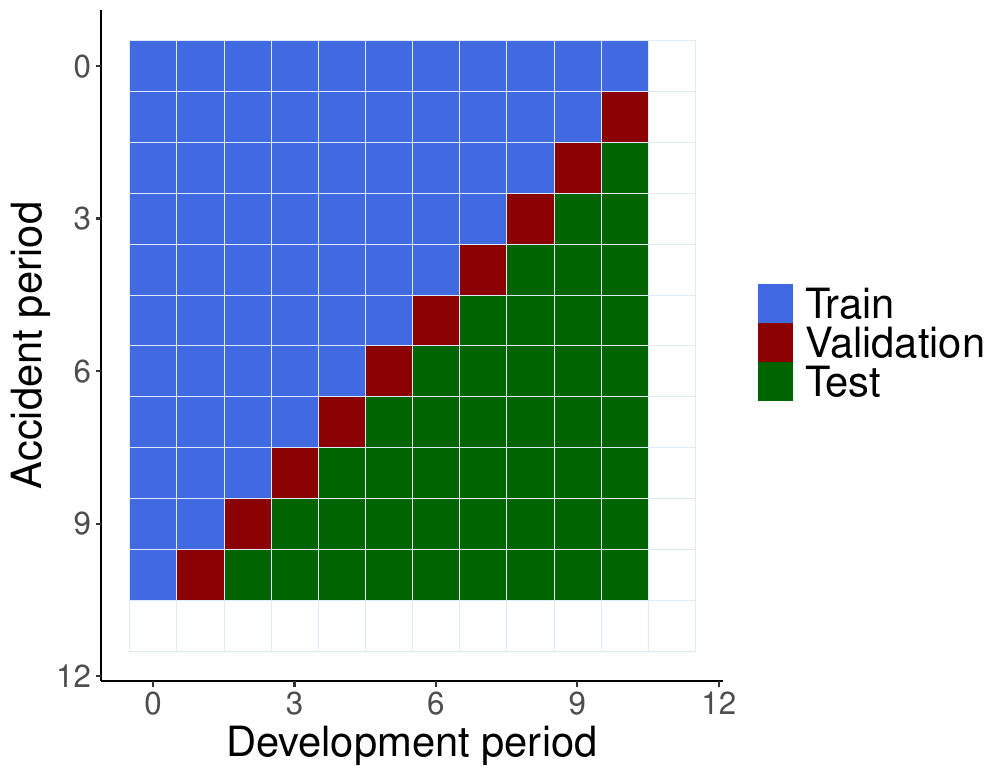}
        \caption{\label{sfig:valibonaic} The triangle is divided into training set, validation set and test set.}
        
    \end{subfigure}

\end{figure}

The data split for the application in Section \ref{sec:naic}, is shown in Figure \ref{fig:naicbo}.

\section{Claims reserving with clmplus}
\label{appendix:clmpluscode}

In this appendix we show the \texttt{clmplus} code implementation of some of the models we showed in the previous Sections.
Version $1.0.0$ of the \texttt{clmplus} package can be installed using the \texttt{install.packages} function

\begin{lstlisting}
install.packages("clmplus")
\end{lstlisting}

{\color{black} Additional reading material can be found on \href{https://gpitt71.github.io/clmplus/}{the package webpage}. }




After loading the \texttt{clmplus} and \texttt{ChainLadder} packages, the dataset of cumulative claim payments is transformed into a \texttt{AggregateDataPP} object that extracts from the data the information to model the development factors as we discussed in this work. 

\begin{lstlisting}
library(clmplus)
library(ChainLadder)

data ("AutoBI")
dataset = AutoBI$AutoBIPaid
datapp = AggregateDataPP(cumulative.payments.triangle = dataset)
\end{lstlisting}

We can see once more the parallel between non-life insurance and mortality modeling by showing the Lexis representation of the triangle of incremental payments in \texttt{AutoBI}.
In the triangle of incremental payments each diagonal represents the calendar year (period), the rows represent the accident year (cohort) and the columns are development year (age).

\begin{lstlisting}
datapp$incremental.payments.triangle
\end{lstlisting}

\begin{lstlisting}
     [,1] [,2] [,3] [,4] [,5] [,6] [,7] [,8]
[1,] 1904 3494 2098 1386  830  359  128   57
[2,] 2235 4026 2430 1752  903  408  277   NA
[3,] 2441 4907 3314 1993 1093  487   NA   NA
[4,] 2503 5670 3637 2366 1207   NA   NA   NA
[5,] 2838 5874 4016 2550   NA   NA   NA   NA
[6,] 2405 5453 3913   NA   NA   NA   NA   NA
[7,] 2759 6423   NA   NA   NA   NA   NA   NA
[8,] 2801   NA   NA   NA   NA   NA   NA   NA
\end{lstlisting}

In the Lexis representation that \texttt{clmplus} uses for modeling each diagonal represents the cohorts (accident year), the rows represent the age (development year) and the columns are period (calendar year).

\begin{lstlisting}
datapp$occurrance
\end{lstlisting}

\begin{lstlisting}
     [,1] [,2] [,3] [,4] [,5] [,6] [,7] [,8]
[1,] 1904 2235 2441 2503 2838 2405 2759 2801
[2,]   NA 3494 4026 4907 5670 5874 5453 6423
[3,]   NA   NA 2098 2430 3314 3637 4016 3913
[4,]   NA   NA   NA 1386 1752 1993 2366 2550
[5,]   NA   NA   NA   NA  830  903 1093 1207
[6,]   NA   NA   NA   NA   NA  359  408  487
[7,]   NA   NA   NA   NA   NA   NA  128  277
[8,]   NA   NA   NA   NA   NA   NA   NA   57
\end{lstlisting}



Starting from the \texttt{datapp} object we can replicate the chain-ladder reserve with an age model. 

\begin{lstlisting}
a.model.fit=clmplus(datapp,
                    hazard.model = "a")
\end{lstlisting}

\begin{lstlisting}
a.model<-predict(a.model.fit,
                hazard.model = "a")
# clmplus reserve (age model)
sum(a.model$reserve)
#31754.43
\end{lstlisting}

In order to show the comparison, we compute the ultimate cost for the Mack chain-ladder model using the \texttt{MackChainLadder} function from the \texttt{ChainLadder} library. 
By subtracting the diagonal cumulative amount we find the chain-ladder reserve.

\begin{lstlisting}
#Mack ultimate cost
mck.ultimate=ChainLadder::MackChainLadder(dataset)$FullTriangle[,8]

#Mack reserve
sum(mck.ultimate - rtt$diagonal)
#31754.43

\end{lstlisting}

In a similar fashion to the age-model (\texttt{a.model}), we can define the age-cohort model (\texttt{ac.model}), the age-cohort model (\texttt{ap.model}), and age-period-cohort model (\texttt{apc.model}).

\begin{lstlisting}
ac.model.fit=clmplus(datapp,
                 hazard.model = "ac")
ac.model=predict(ac.model.fit,
                 gk.fc.model = 'a',
                 gk.order = c(1,1,0))

# clmplus reserve (age-cohort model)
sum(ac.model$reserve)
#38126.05

ap.model.fit=clmplus(datapp,
                 hazard.model = "ap")

ap.model<- predict(ap.model.fit,
                 ckj.fc.model = 'a',
                 ckj.order = c(0,1,0))

# clmplus reserve (age-period model)
sum(ap.model$reserve)
#37375.01

apc.model.fit=clmplus(datapp,
                  hazard.model = "apc")
apc.model<-predict(apc.model.fit,
                  gk.fc.model = 'a',
                  ckj.fc.model = 'a',
                  gk.order = c(1,1,0),
                  ckj.order = c(0,1,0))
# clmplus reserve (age-period-cohort model)
sum(apc.model$reserve)
#38498.54

\end{lstlisting}

The models residuals in Figure \ref{fig:modelsresiduals} can be easily obtained with the \texttt{plotresiduals} function. We show an example for the age model to obtain Figure \ref{fig:resa}.

\begin{lstlisting}
plot(a.model.fit)

\end{lstlisting}

The \texttt{clmplus} age model effect in Figure \ref{fig:clmplus} can be inspected with the \texttt{plot} function:

\begin{lstlisting}
plot(a.model)
\end{lstlisting}

For those models that require to extrapolate a period effect it is possible to inspect the fitted and extrapolated parameters with \texttt{clmplus}. Fitted and extrapolated effects for the age-period-cohort model:

\begin{lstlisting}
plot(apc.model)
\end{lstlisting}

\begin{figure}[H]
        \centering
        \includegraphics[width=15cm]{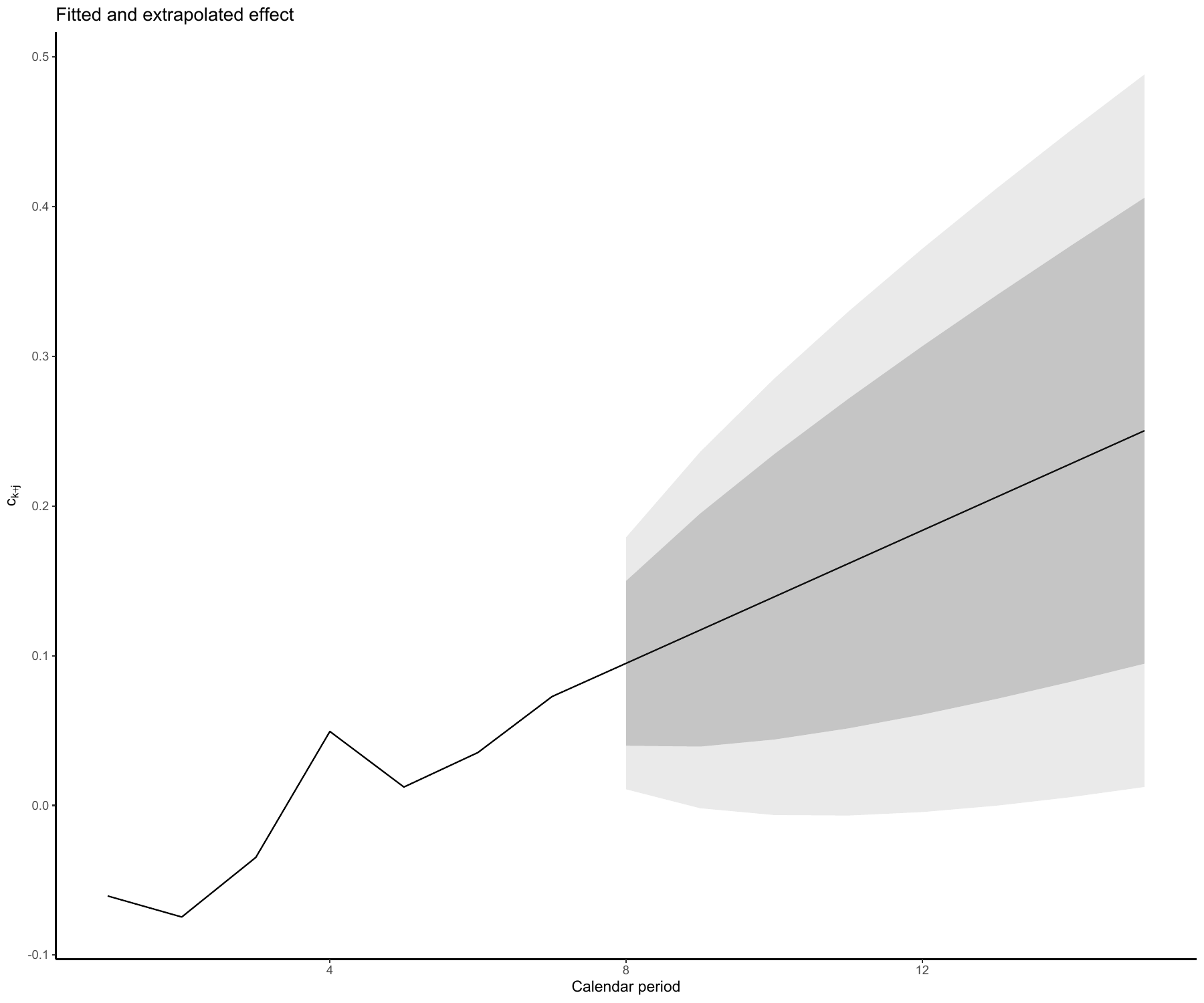}
\end{figure}

\section{Triangles tested}
\label{appendix:triangleslist}

  \begin{table}[H]
\centering
\begin{tabular}{rlr}
  \hline
&Dataset & Data source (package) \\ 
  \hline
1&GenIns& \texttt{ChainLadder} \\ 
2&sifa.mod & \texttt{clmplus} \\ 
3&  sifa.gtpl & \texttt{clmplus} \\ 
 4& sifa.mtpl & \texttt{clmplus} \\ 
 5& amases.gtpl & \texttt{clmplus} \\ 
 6& amases.mod & \texttt{clmplus} \\ 
 7& amases.mtpl & \texttt{clmplus} \\ 
 8& bz & \texttt{apc} \\
 9 & ta & \texttt{apc} \\
10&  xl & \texttt{apc} \\
 11& vnj & \texttt{apc} \\
 12&  abc& \texttt{ChainLadder} \\ 
 13&  autoC& \texttt{ChainLadder} \\ 
 14&  autoP & \texttt{ChainLadder} \\ 
15&   autoBI & \texttt{ChainLadder} \\ 
16&   mclpaid& \texttt{ChainLadder} \\ 
17&   medmal& \texttt{ChainLadder} \\ 
 18&  mortgage& \texttt{ChainLadder} \\ 
19&   mw08& \texttt{ChainLadder} \\ 
20&   mw14& \texttt{ChainLadder} \\ 
21&   ukmotor & \texttt{ChainLadder} \\ 
22&   usapaid& \texttt{ChainLadder} \\ 
 23&  aus1& \texttt{CASdatasets} \\ 
24&   fre4b& \texttt{CASdatasets} \\ 
 25&  swiss1& \texttt{CASdatasets} \\ 
26&   usa1& \texttt{CASdatasets} \\ 
27&   usa2& \texttt{CASdatasets} \\ 
28&   usa3 & \texttt{CASdatasets} \\ 
29&   usa4& \texttt{CASdatasets} \\ 
30&   usa5& \texttt{CASdatasets} \\ 
   \hline
\end{tabular}
\caption{\label{tab:sectionfive} List of the run-off triangles that we used in Section \ref{sec:ModelComparison}.}
\end{table}


\section{Simulation study on environmental changes}

\label{appendix:simulationstudy}

\begin{figure}
\centering
\includegraphics[width=\linewidth]{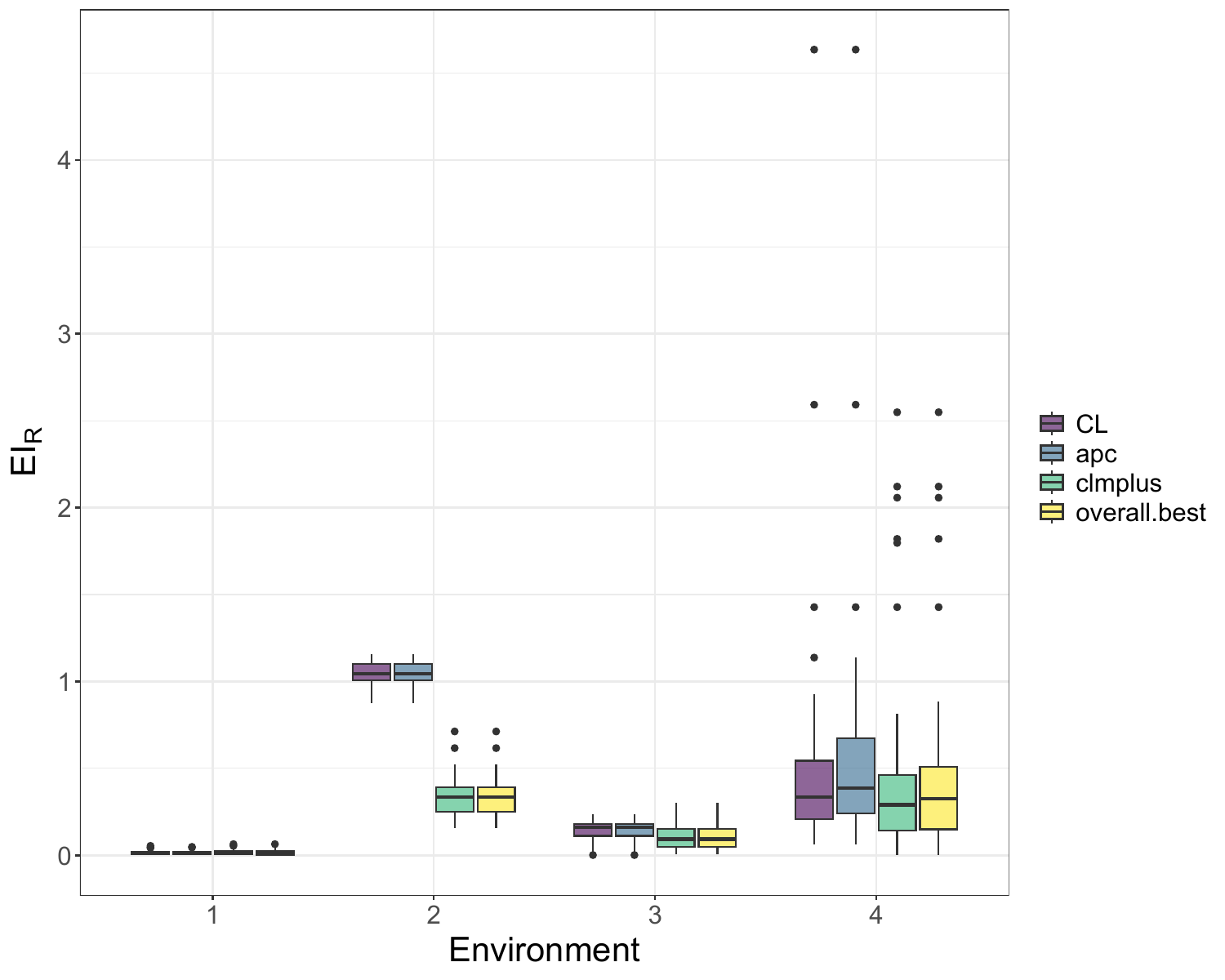}
        \caption{Results on three of the simulated Environments from \protect\cite{almudafer21}. For each model (\texttt{a}, \texttt{ac}, \texttt{ap}, \texttt{apc}), we have three box-plots of the $\text{EI}_\text{R}$ on the data simulated from Environment $1$ (violet), Environment $2$ (blue), Environment $3$ (light blue) and Environment $4$ (yellow).}
        \label{fig:simbxpl}
\end{figure}

{\color{black} In this Section, we apply our models to the simulated scenarios from \citeA{almudafer21}, using the same approach described in Section \ref{ss:environment2}. Before proceeding with our application, we provide a brief description of the four scenarios. A more comprehensive documentation can be found in the original manuscript. In \autoref{fig:edasimulations} for each Environment we extract one simulated data set and show an exploratory data analysis on the evolution of incremental payments in the different development periods for accident periods $10, 20, 30, 40$.

\begin{figure}[H]
    \centering
    \caption{\label{fig:edasimulations} Incremental Claim Amount on Environment $1$ (\autoref{fig:simeda1}), Environment $2$ (\autoref{fig:simeda2}), Environment $3$ (\autoref{fig:simeda3}), Environment $4$ (\autoref{fig:simeda1}) for one simulation for each accident period in Accident Quarters (AQ) $1, 10, 20, 30, 40$. The amounts are displayed in millions.}
    \begin{subfigure}[t]{0.49\textwidth}
        \centering\includegraphics[width=\linewidth]{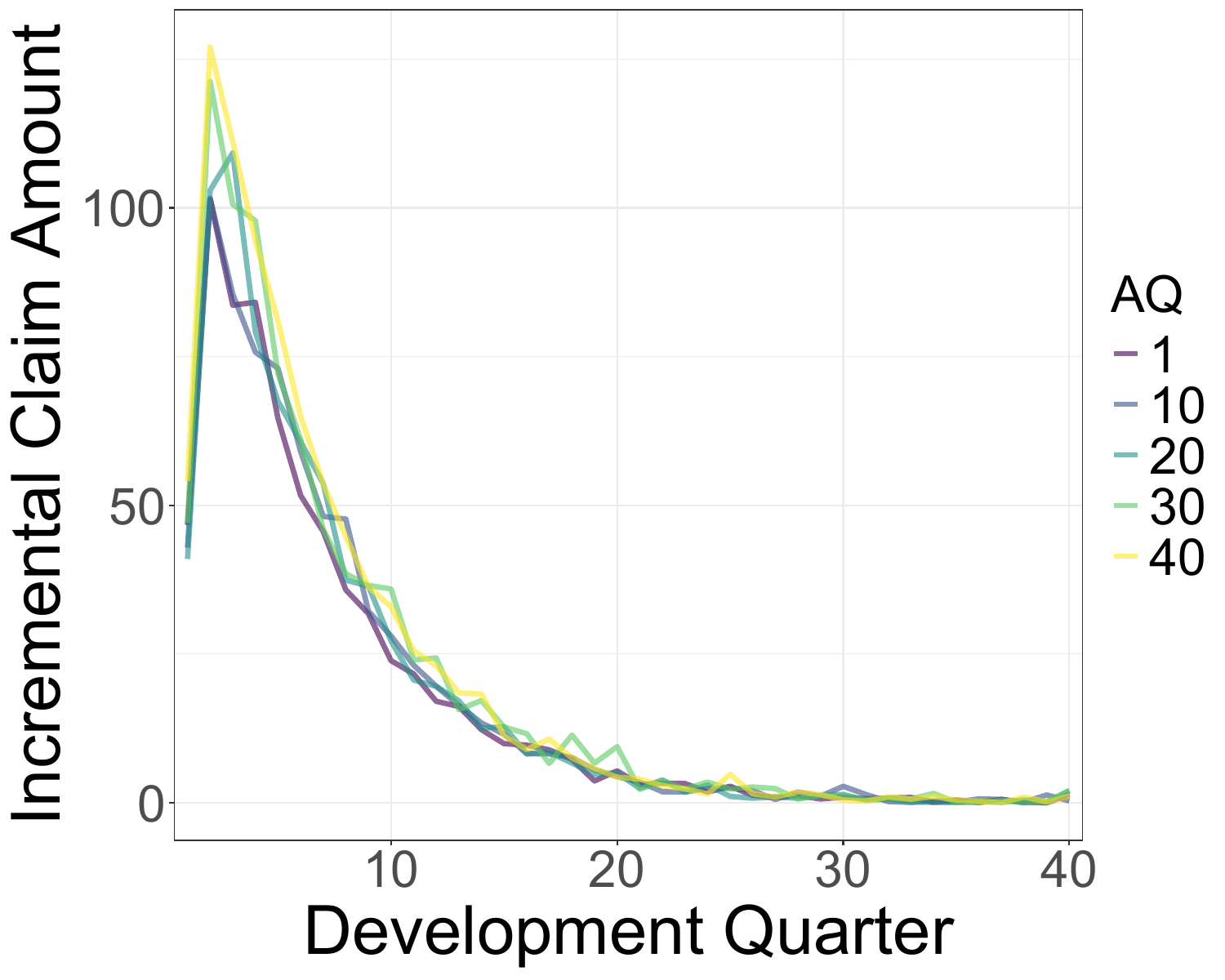}
        \caption{\label{fig:simeda1} }

    \end{subfigure}
    \hfill
    \begin{subfigure}[t]{0.49\linewidth}
        \centering\includegraphics[width=\linewidth]{figures/simeda2.pdf}
        \caption{\label{fig:simeda2} }
        
    \end{subfigure}
    \hfill
    \begin{subfigure}[t]{0.49\linewidth}
        \centering\includegraphics[width=\linewidth]{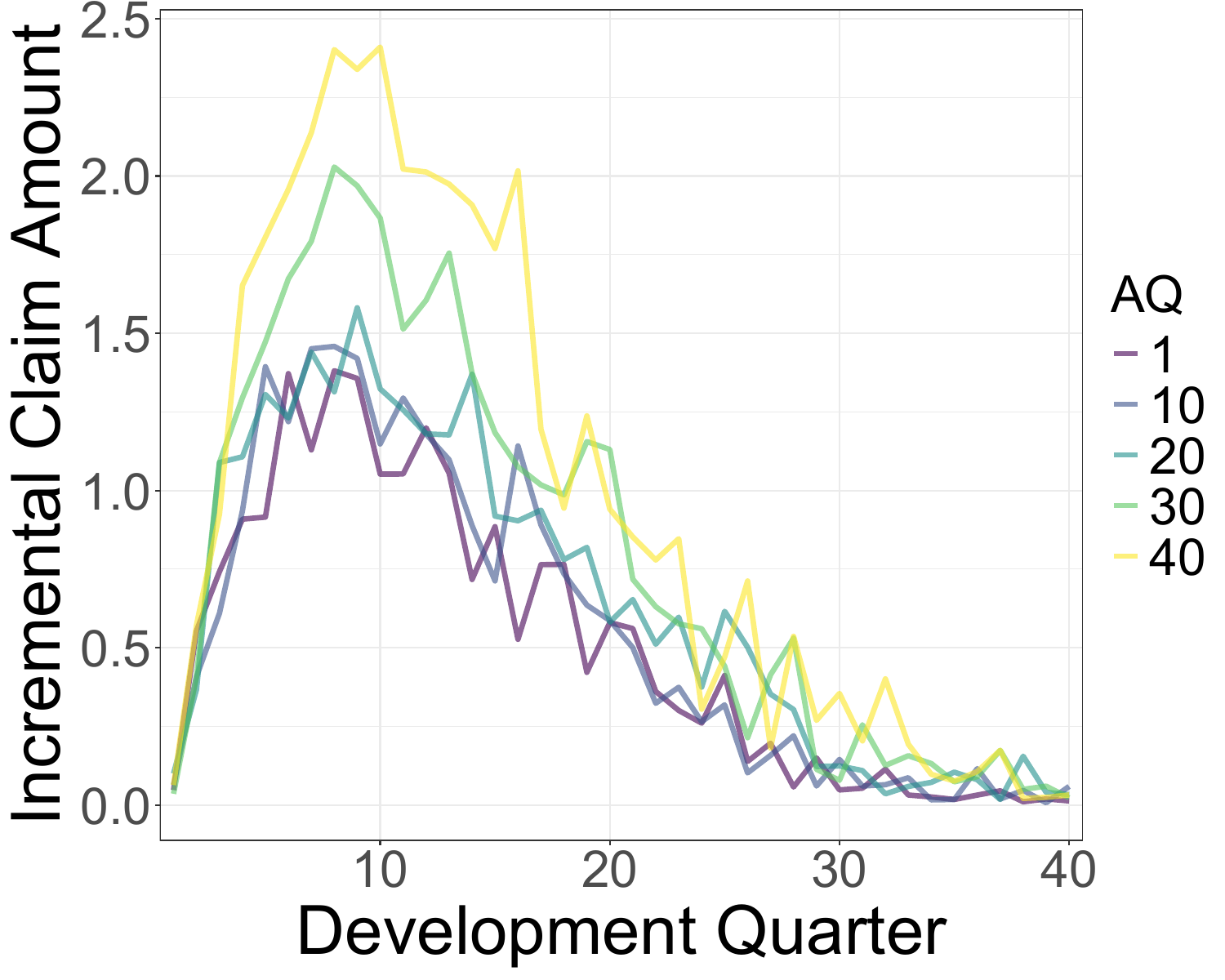}
        \caption{\label{fig:simeda3} }
        
    \end{subfigure}
    \hfill
    \begin{subfigure}[t]{0.49\textwidth}
        \centering\includegraphics[width=\linewidth]{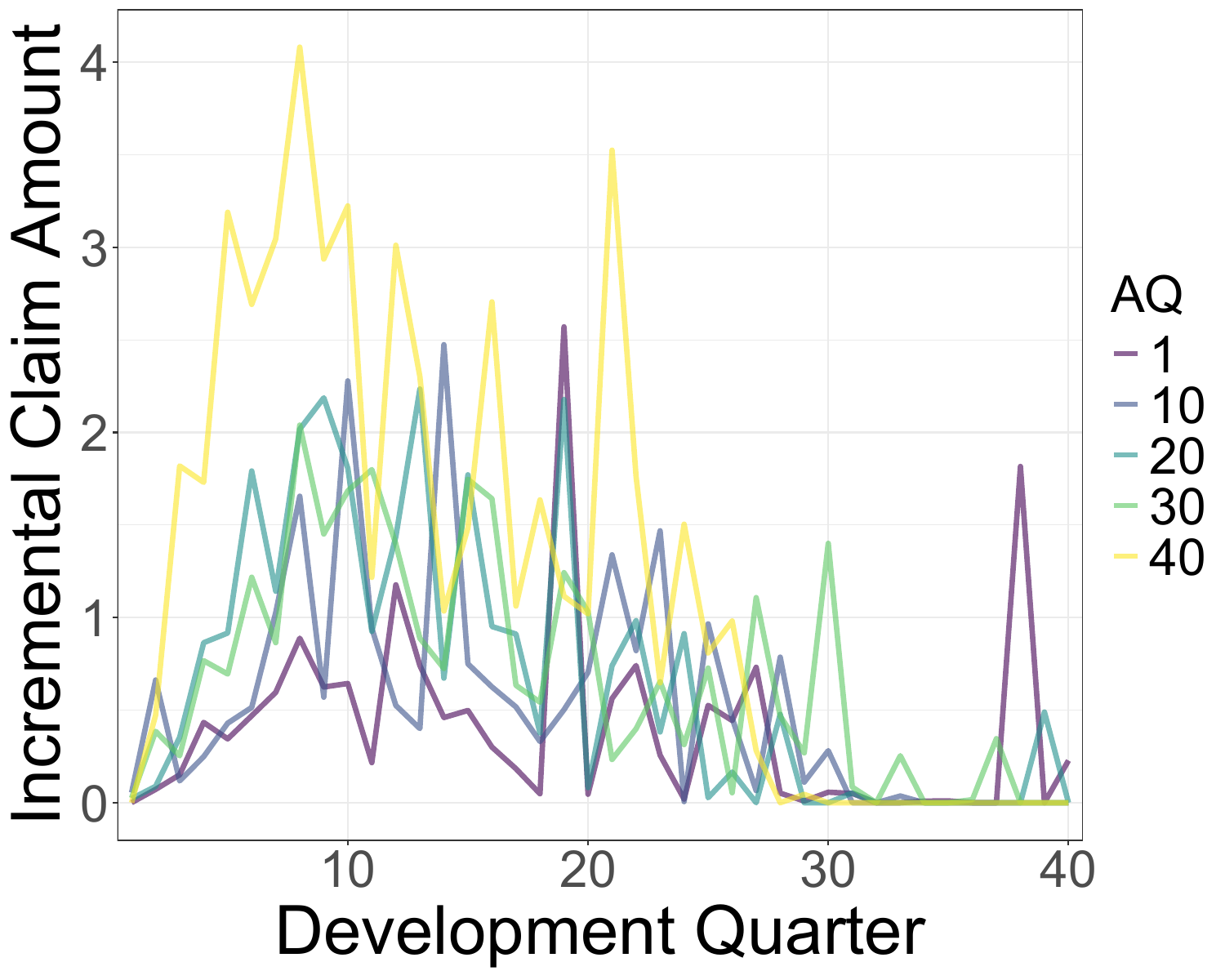}
        \caption{\label{fig:simeda4} }
        
    \end{subfigure}
\end{figure}

Environment $1$ (\autoref{fig:simeda1}) constitutes a baseline scenario of simple, short tail claims where the chain-ladder is expected to work well. Conversely, Environment $2$ (\autoref{fig:simeda2}) and Environment $3$ (\autoref{fig:simeda3}) include environmental changes. In Environment $2$, there is an accident period trend in terms of increase in the settlement speed for claims from later accident periods. In Environment $3$ there is a calendar period effect in the form of  inflation on the claim amounts. We observe that the type of environmental changes occurring in Environments $2$ and $3$ are also identified in \citeA{okine22} as a typical case where traditional reserving methods like the chain ladder might fail. Environment $4$ (\autoref{fig:simeda4}) includes a super-imposed inflation effect and a complex dependency structure between reporting delays, claims severity and settlement delay.
\newline

In our application, for each Environment ($1$, $2$, $3$ and $4$) we download from the \texttt{GitHub} folder \citeA{agiLabReservingMDN} the data that \citeA{almudafer21} used in their paper data application. For each of the Environments the authors simulated $50$ reserving data sets. On each of these copies, we repeat the analysis in terms of $\text{EI}_{\text{R}}$ that we described in Section \ref{ss:environment2}. 

For validation and testing, we use the same approach illustrated in Figure \ref{fig:naicbo}. In particular, we use the data from calendar period $k+j=m$ for validation and the data from calendar periods $k+j>m$ for testing. The true reserve is available from the simulations. 

We summarize the results for each Environment in the box-plot in Figure \ref{fig:simbxpl}. Each data point in the Figure, is an $\text{EI}_{\text{R}}$ measurement for one simulation.

The results show that in Environment $1$, on average the age-model (the chain-ladder) performs similarly to the three model families. The results for Environment $2$ are reported from Section \ref{ss:environment2}. Similar conclusions hold for Environment $3$, where a calendar period effect is present. In particular, we see that using the complete set of models we can improve the predictive performance of our models in terms of $\text{EI}_{\text{R}}$. In Environment $4$, where the data generating process includes monetary inflation and a complicated simulation mechanism, all of the model families seem to have a modest $\text{EI}_{\text{R}}$ performance. However, \texttt{clmplus} seem to provide better results than the \texttt{apc} family.

Interestingly, having available the complete set of models in the \texttt{overall.best} family, allows to improve the $\text{EI}_{\text{R}}$ accuracy, indicating that with our validation approach we are able to select the preferable model.

The frequency of the model chosen in each scenario as overall best model is shown in Table \ref{tab:setcchoices}.

}

\begin{table}[ht]
\centering
\begin{tabular}{lrr}
  \hline
    Models (Family)& Environment & Set (c) Choice \\ 
  \hline
   \texttt{ac} (\texttt{apc})/\texttt{a} (\texttt{clmplus}) & \multirow{5}{*}{1} &  12 \\ 
   \texttt{apc} (\texttt{apc}) &  &  16 \\ 
   \texttt{ac} (\texttt{clmplus}) &  &   7 \\ 
   \texttt{ap} (\texttt{clmplus}) &  &   8 \\ 
   \texttt{apc} (\texttt{clmplus}) &  &   7 \\ 
  \hline
   \texttt{ac} (\texttt{clmplus}) & \multirow{3}{*}{2} &   1 \\ 
   \texttt{ap} (\texttt{clmplus}) &  &  30 \\ 
   \texttt{apc} (\texttt{clmplus}) &  &  19 \\ 
  \hline
   \texttt{ac} (\texttt{apc})/\texttt{a} (\texttt{clmplus}) & \multirow{4}{*}{3} &   8 \\ 
   \texttt{ac} (\texttt{clmplus}) &  &  16 \\ 
   \texttt{ap} (\texttt{clmplus}) &  &  15 \\ 
   \texttt{apc} (\texttt{clmplus}) &  &  11 \\ 
  \hline
   \texttt{ac} (\texttt{apc})/\texttt{a} (\texttt{clmplus}) & \multirow{5}{*}{4} &  14 \\ 
   \texttt{apc} (\texttt{apc}) &  &   5 \\ 
   \texttt{ac} (\texttt{clmplus}) &  &  17 \\ 
   \texttt{ap} (\texttt{clmplus}) &  &   7 \\ 
   \texttt{apc} (\texttt{clmplus}) &  &   7 \\ 
  \hline
\end{tabular}
\caption{\label{tab:setcchoices} For each scenario from {\protect \citeA{almudafer21}}, we present the selection frequency of each model based on our validation approach.}
\end{table}

\end{document}